\documentclass[openacc]{rstransa}

\usepackage{natbib}
\usepackage{graphicx}
\usepackage{amsmath}

\newcommand\mancha{\textsc{Mancha3D~}}
\newcommand\batt{\textsc{batt}}
\newcommand\ambi{\textsc{ambi}}
\newcommand\ambihall{\textsc{ambihall}}

\def\specchar#1{{\sc #1}}
\def\CaII{\mbox{Ca\,\specchar{ii}}}

\begin{document}

\title{Influence of ambipolar and Hall effects on vorticity in 3D simulations of magneto-convection}

\author{E. Khomenko$^{1,2}$, M. Collados$^{1,2}$, N. Vitas$^{1,2}$, P. A. Gonz\'alez-Morales$^{1,2}$}

\address{$^{1}$Instituto de Astrof\'{\i}sica de Canarias, 38205 La Laguna, Tenerife, Spain\\
$^{2}$Departamento de Astrof\'{\i}sica, Universidad de La Laguna, 38205, La Laguna, Tenerife, Spain}

\subject{sun, atmosphere, waves, simulations}


\corres{Elena Khomenko\\
\email{khomenko@iac.es}}

\begin{abstract}
This paper presents the results of the analysis of 3D simulations of solar magneto-convection that include the joint action of the ambipolar diffusion and the Hall effect. Three simulation-runs are compared: one including both ambipolar diffusion and Hall effect; one including only ambipolar diffusion; and one without any of these two effects. The magnetic field is amplified from initial field to saturation level by the action of turbulent local dynamo. In each of these cases, we study 2 hours of simulated solar time after the local dynamo reaches the saturation regime. We analyze the power spectra of vorticity, of magnetic field fluctuations and of the different components of the magnetic Poynting flux responsible for the transport of vertical or horizontal perturbations. Our preliminary results show that the ambipolar diffusion produces a strong reduction of vorticity in the upper chromospheric layers and that it dissipates the vortical perturbations converting them into thermal energy. The Hall effect acts in the opposite way, strongly enhancing the vorticity.  When the Hall effect is included, the magnetic field in the simulations becomes, on average, more vertical and long-lived flux tube-like structures are produced. We trace a single magnetic structure to study its evolution pattern and the magnetic field intensification, and their possible relation to the Hall effect. 
\end{abstract}

\begin{fmtext}
\end{fmtext}


\maketitle

\section{Introduction}


Vortical motions are an intrinsic part of the solar atmosphere dynamics. These motions are observed from extremely small scales, down to sub-arcseconds in  photospheric granulation \cite{Bonet2008, Vargas2011}, and up to scales seen in giant tornados in solar prominences \cite{OrozcoSuarez2012} (see Tziotziou et al. \cite{Tziotziou2020} for a review). In the chromosphere,  Wedemeyer-B\"ohm et al. \cite{Wedemeyer2009} discovered for the first time conspicuous swirling motions in the observations of the \CaII\ 8532 \AA\ line done with CRISP instrument at the SST. The size of the observed swirls was typically about 2 arcsec with smaller fragments down to 0.2 arcsec consisting of arcs and spirals. These swirls were observed to be co-spatial with photospheric bright points moving with respect to each other. Wedemeyer-B\"ohm et al. \cite{Wedemeyer2009} suggested that spiral motions,  guided by the magnetic field, propagate along magnetic structures from the photosphere to the chromosphere. Later, Wedemeyer-B\"ohm et al. \cite{Wedemeyer2012} reported on observations of the extension of swirls from the photosphere and chromosphere into the transition region and low corona, and associated the swirls with rotating magnetic field structures. It was suggested that magnetic swirls serve as energy channels allowing the energy transport from the photosphere to the corona and that they are important contributors to the chromospheric energy budget \cite{Wedemeyer2012, Park+etal2016, Yadav2020}. 

The typical lifetimes of swirls detected in observations are of the order of tens of minutes \cite{Wedemeyer2012, Shetye2019}. Oscillatory-like behaviour is frequent for these structures with periods of about 3 minutes, see Tziotziou et al. \cite{Tziotziou2019b}. There is evidence that incompressible waves can be excited by the vortex motions of a strong magnetic flux concentration in the photosphere and then these waves detected in the chromosphere, Morton et al. \cite{Morton2013}. Non-magnetic swirls usually do not extend much in height. The magnetic swirls detected simultaneously at several layers usually appear above network magnetic field concentrations \cite{Wedemeyer2012, Shetye2019}. Nevertheless, an extremely long-lived feature, of at least 1.7 hours, was detected in quiet Sun observations by Tziotziou et al. \cite{Tziotziou2018}. The authors could not find evidence of magnetically driven flows and no association with photospheric magnetic bright points was observed. The vortex flow resembled a small-scale tornado with internal substructure and had a relatively large radius of about 3 arcseconds. 


Significant modelling efforts have been made in an attempt to understand the physics of swirling motions in the solar atmosphere. They range from numerical experiments where vortices are generated in idealized conditions, such as isolated flux tubes, controlled driver properties \cite{Fedun+etal2011, Chmielewski014, Zaqarashvili2015, Shelyag+etal2016, Snow+etal2018, Khomenko+Cally2019}, to more realistic models where the vortical motions are naturally produced by magneto-convection \cite{Nordlund+Stein2001, Stein+Nordlund2001, Carlsson2010, Moll2011, Shelyag2011, Shelyag2012, Moll2012, Kitiashvili2012, Kitiashvili2013, Wedemeyer2014, Liu2019, Yadav2020}. It has been shown that the nature of magnetic and non-magntic vortices is intrinsically different, see Shelyag et al. \cite{Shelyag2011}. In the non-magnetic case, vorticity is generated by the baroclinic term in the governing equation of vorticity. This term is proportional to the cross-product of pressure and density gradients, and thus enhanced in intergranular lanes.  In the magnetic case, magnetic tension dominates over baroclinicity \cite{Shelyag2011}. In the work by Moll et al. \cite{Moll2012}, vertically orientated vortices extending over the entire height of the simulated photosphere were found in the magnetic case. Both magnetic and non-magnetic vortex structures were found to be closely related to local heating, see Moll et al. \cite{Moll2011}. In the magnetized case, Wedemeyer \& Steiner \cite{Wedemeyer2014} found two vortex flow systems  stacked on top of each other. The lower vortex that extends from the convection zone to the low photosphere is an intergranular vortex flow. Once a magnetic field structure is co-located with this intergranular vortex flow, the rotation is extended into the upper atmospheric layers. Therefore, simulations of strong magnetic swirls require extension to sufficiently high layers and inclusion of the magnetic field  \cite{Wedemeyer2014}. Penetration of the vortex tube causes significant thermodynamic changes in the low chromosphere, see Kitiashvili et al. \cite{Kitiashvili2013}. While the pressure drops in the vortex core, the mass continues to accumulate around it, creating significant plasma gradients. The temperature in the vortex core is lower relative to the surrounding in the photosphere, and higher at about \hbox{600 km} above the photosphere \cite{Kitiashvili2012, Kitiashvili2013}. The continuous build up of the energy may lead to prominence eruptions, as in Kitiashvili et al. \cite{Kitiashvili2013}. Plasma is accelerated upwards and outwards by the centrifugal force during the vortex, and significant amount of the Poynting flux to the corona is generated in the form of torsional Alfv\'en waves  \cite{Wedemeyer2012}.

None of the simulations of vortex formation done so far include non-ideal plasma effects, such as ambipolar diffusion due to the presence of neutrals in the solar plasma, or the Hall effect (which is also modified in the presence of neutrals). The ambipolar diffusion is considered as one of the promising mechanisms to take into account in the models of chromospheric heating \cite{Khomenko+Collados2012, MartinezSykora+etal2012, MartinezSykora2017, Ballester+etal2018, Khomenko+etal2018, Nobrega-Siverio+etal2020}. It helps to dissipate non-compressible perturbations, such as those associated to Alfv\'en waves, and therefore it can impact vorticity as well. The Hall effect is insufficiently studied, although there are theoretical considerations showing that its continuous action helps production of Alfv\'en waves in the chromosphere \cite{Cheung+Cameron2012, Cally+Khomenko2015, Gonzalez-Morales+etal2019}, or the formation of strong magnetic field structures \cite{Khodachenko+Zaitsev2002}. Hall effect has been included in 2.5D magneto-convection simulations, by Martinez-Sykora et al. \cite{MartinezSykora+etal2012} and Cheung \& Cameron \cite{Cheung+Cameron2012}. In the simulations of the umbra magneto-convection, Cheung \& Cameron \cite{Cheung+Cameron2012} describe the generation of a small, out of the 2D plane, component of the velocity (of the order of 100 m/s) and magnetic field (of about 5 G strength, compared to the kG field in the umbra), which is then advected by convection. The Hall effect is an intrinsically 3D effect, because it creates velocities in the direction perpendicular to both, the magnetic field, and the (arbitrary) initial perturbation, $\mathbf{v}^{\rm Hall}\sim \mathbf{J}$.  Therefore, its full action can only be addressed in 3D experiments. 

The aim of the current work is to start filling this gap and to provide an initial study of the influence of non-ideal effects on vorticity in realistic 3D simulations of small-scale solar dynamo in the saturated regime. We continue the study from Khomenko et al. \cite{Khomenko+etal2018} (Paper I) and Gonz\'alez-Morales et al. \cite{2020A&A...XXXX..XXX} (Paper II) on the action of the ambipolar diffusion and Hall effect in 3D models of solar magneto-convection, with a special emphasis on the vorticity generation and dissipation, and on modifications of the average magnetic structure of the atmosphere due to the Hall effect.

\begin{figure}  
\begin{center}
\includegraphics[width = 13cm]{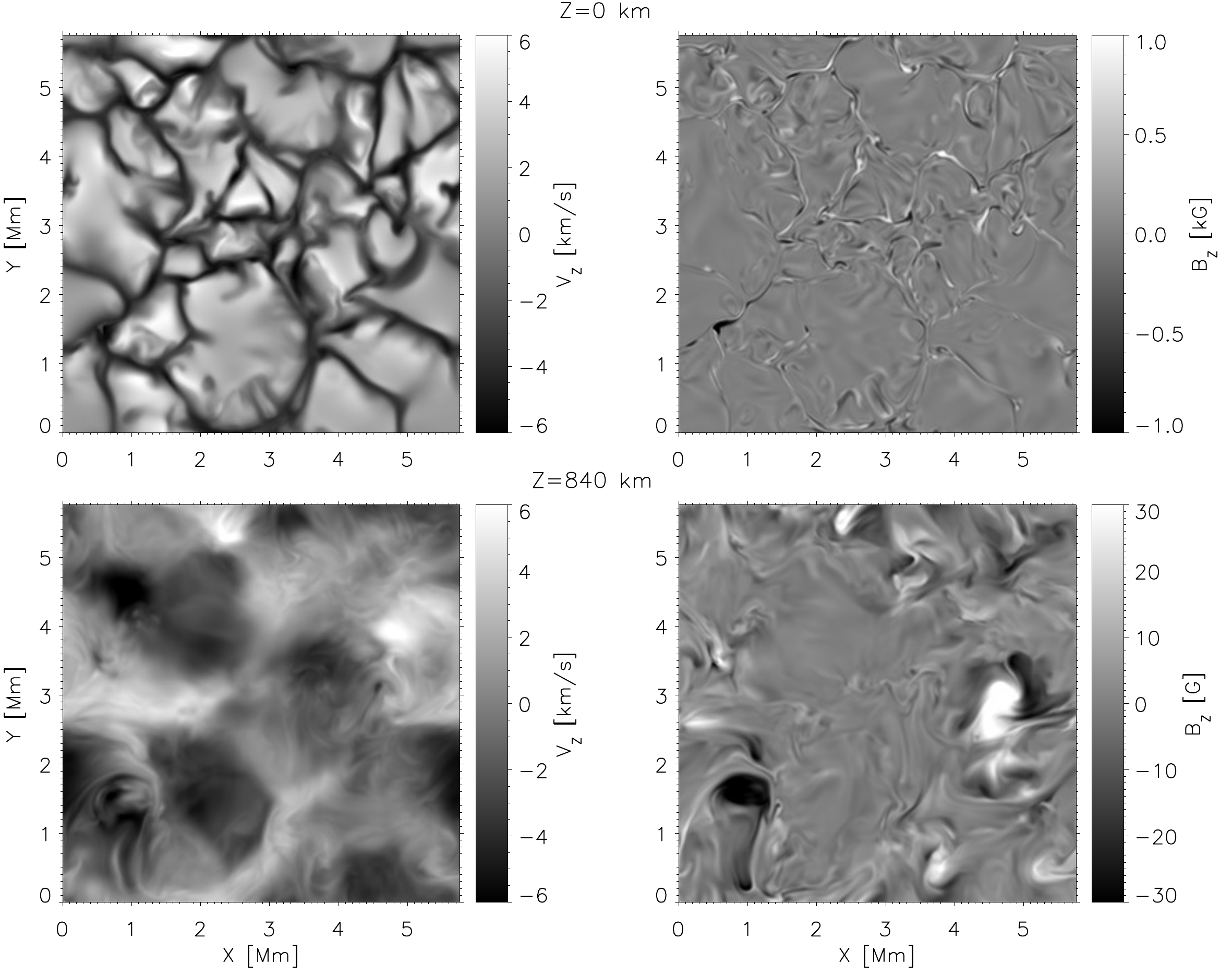}
\end{center}
\caption{Horizontal slices through the simulation domain of the \ambihall\ simulation at t=34 min after the start of the analyzed series at height 0 km (top) and 840 km (bottom). Snapshots on the left show the vertical component of the velocity; snapshots on the right show the vertical component of the magnetic field. }
\label{fig:visual}
\end{figure}

\begin{figure}  
\begin{center}
\includegraphics[width = 13cm]{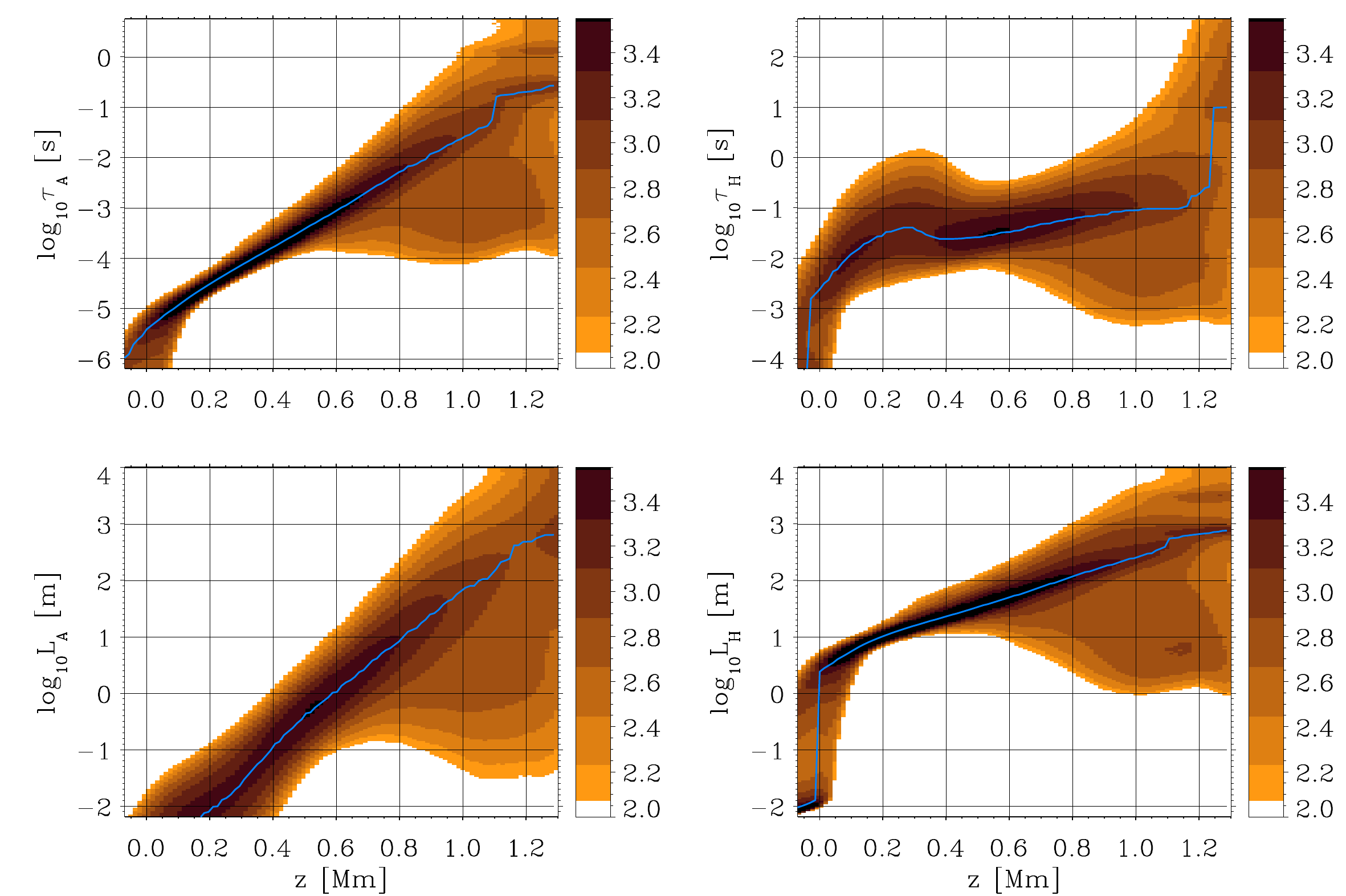}
\end{center}
\caption{Two dimensional histograms showing the number of occurrences of a given value of a parameter as a function of height in the \ambihall\ run, with darker colours indicating larger probability in log$_{10}$ scale. The blue lines show the median value of the distribution. Top left: temporal scale of the ambipolar diffusion, $\tau_A$; top right: temporal scale of the Hall effect, $\tau_H$; bottom left: spatial scale of the ambipolar diffusion, $L_A$; bottom right: spatial scape of the Hall effect, $L_H$. Scales on the vertical axes are shown in log$_{10}$ units.}
\label{fig:scales}
\end{figure}

\section{Method}

The simulations that we analyze in this paper were previously used in Paper II \cite{2020A&A...XXXX..XXX}, and represent a continuation of the battery-excited small-scale solar dynamo series from Khomenko et al. \cite{Khomenko+etal2017} and Paper I \cite{Khomenko+etal2018}. The simulations are obtained with the \mancha code \cite{Khomenko+Collados2006, Felipe+etal2010, Gonzalez-Morales+etal2018} and are fully described in the publications mentioned above.  The simulations take into account partial ionization of the solar atmosphere through the single-fluid approximation using the generalized induction equation, and including the corresponding terms due to neutrals in the total energy equation. The following physical set of equations, written for perturbations, is solved by the \mancha code for these simulations,
\begin{eqnarray}\label{eq:continuity} 
\frac{\partial \rho_1 }{ \partial t} + \mathbf{\nabla} \cdot \left( \rho\mathbf{v} \right) =  0,
\end{eqnarray}
\begin{eqnarray} \label{eq:momentum} 
\frac{\partial \rho\mathbf{v} }{\partial t} + \mathbf{\nabla}\cdot \left[\rho\mathbf{v} \mathbf{v} + \left(p_1+ \frac{\mathbf{B}_1^2 + 2\mathbf{B}_1 \cdot \mathbf{B}_0}{2 \mu_0} \right) \mathbf{I} \right] +\mathbf{\nabla}\cdot \left[\frac{1}{\mu_0} \left(\mathbf{B}_0 \mathbf{B}_1-\mathbf{B}_1 \mathbf{B}_0-\mathbf{B}_1 \mathbf{B}_1 \right) \right] = \rho_1 \mathbf{g} ,
\end{eqnarray}
\begin{eqnarray} \label{eq:induction}
\frac{\partial \mathbf{B}_1}{\partial t}  =  \mathbf{\nabla}\times \left[\mathbf{v}\times \mathbf{B} \right] +  \mathbf{\nabla}\times \left[\frac{\mathbf{\nabla}p_e}{e n_e} -\eta_H\frac{(\mathbf{J} \times \mathbf{B})}{|B|} + \eta_A\frac{\left(\mathbf{J} \times \mathbf{B} \right) \times \mathbf{B} }{|B|^2}   \right],
\end{eqnarray}
\begin{eqnarray} \label{eq:energy}
\frac{\partial e_1}{\partial t} + \nabla \cdot \left[ \mathbf{v}\left(e + p + \frac{|\mathbf{B}|^2}{2 \mu_0} \right) - \frac{\mathbf{B}(\mathbf{v} \cdot \mathbf{B}) }{\mu_0} \right]   = \rho \left(\mathbf{g} \cdot \mathbf{v}\right)  +\mathbf{\nabla}\cdot \left[ \eta_A \frac{\mathbf{B} \times \mathbf{J_\perp} }{\mu_0}   + \frac{\mathbf{\nabla}p_e \times \mathbf{B}} {e n_e \mu_0}    \right] + Q_{\rm R} . 
\end{eqnarray}
The meaning of all variables is standard. The right-hand-side of the generalized induction equation (\ref{eq:induction}) contains the battery, Hall and ambipolar terms; the total energy equation (\ref{eq:energy}) contains the counterparts of the battery and ambipolar terms. The Hall term does not affect the conservation of the total energy.  The ambipolar and Hall coefficients are computed as,
\begin{equation} \label{eq:etaa}
\eta_A =\frac{\xi_n^2|B|^2}{\alpha_n}; \,\,\, \eta_H=\frac{|B|}{en_e},
\end{equation}
in units of $[ml^3/tq^2]$. Here $\xi_n=\rho_n/\rho$ is the neutral fraction, and $\alpha_n$ is the collisional parameter, which includes the electron-ion, and ion-neutral collisions, see Ballester et al.  \citep{Ballester+etal2018}. The electron pressure, $p_e$ and the electron number density, $n_e$, are computed self-consistently from the rest of thermodynamic variables, assuming instantaneous ionization (Saha equation). This is done through pre-computed Equation of State (EOS) tables based on the solar chemical mixture given by Anders \& Grevesse \cite{1989GeCoA..53..197A}. The radiative transfer losses are computed using grey opacity and assuming Local Thermodynamic Equilibrium (LTE).

The variables with subscript $0$ in Eqs. (\ref{eq:continuity}--\ref{eq:energy}) stand for initial model atmosphere variables, and these variables must fulfil the condition of the (magneto)hydrostatic equilibrium (MHS), as described in Paper I. In the current simulations, the $B_0=0$ was initially zero. The perturbation in magnetic field, $B_1$ was initiated through the battery term in the simulations described in Khomenko et al. \cite{Khomenko+etal2017}. The amplitude of the initial seed field was of the order of 10$^{-6}$ G. After seeding the field, it was amplified due to the dynamo action reaching the saturated dynamo regime. Here we  analyze 2 hours of simulated solar time in the saturated dynamo regime. The mean field strength at photospheric level in this regime reached about $|B|=110$ G. 

The grid cell is 20 km $\times$ 20 km in the horizontal plane and 14 km high. The bottom boundary of the domain is located around -0.95 Mm below the solar surface; the upper boundary is located about 1.4 Mm above it. The horizontal size of the domain is $5.8\times5.8$ Mm$^2$. Snapshots are saved every 10 seconds of solar time. The runs analyzed here are labeled as \batt, \ambi\ and \ambihall, following Paper II. The \ambihall\ run includes ambipolar and Hall effects from the generalised induction equation; the \ambi\ run only includes ambipolar effect, and the \batt\ run does not include any of these non-ideal effects and it is used as a reference. The simulations represent quiet areas of the Sun, which are known to cover at least 90\% the solar surface outside active regions at any moment of solar activity cycle. These quiet areas are known to harbour mixed-polarity magnetic elements with strength ranging from hG to kG, and complex structuring \cite{SanchezAlmeida+MartinezGonzales2011}. The magnetic energy contained in solar quiet areas is rather large, as the average field strength, according to the latest studies, can reach as much as 100-130 G \cite{TrujilloBueno2004, Shchukina+TrujilloBueno2011, Danilovic2016}, similar to our simulations. The ambipolar effect, considered in our modeling, provides a way of dissipating this magnetic energy and converting it into the thermal energy, with a potential of being a very effective mechanism of chromospheric heating \cite{Khomenko+Collados2012, MartinezSykora+etal2012, Khomenko+etal2018, Ballester+etal2018}.

Figure \ref{fig:visual} shows an example of simulation snapshots for the \ambihall\ run at a randomly picked time moment at two heights: at the bottom photosphere (top), and at the bottom chromosphere (bottom). The velocity field in the photosphere shows a pattern typical for solar granulation. A comparison to the magnetic field reveals that strong magnetic concentrations with strength up to kG are located in some of the intergranular lanes. The magnetic field is of complex structure, changing the polarity at very small scales. In the chromosphere (bottom), the velocity shows a pattern typical for chromospheric shocks. The structures in magnetic field expand with height, as dictated by the density drop, and occupy larger volume. Two particularly intense structures are seen at the magnetic field image (bottom right). Both of them can be traced down to the photosphere. These magnetic structures keep rotating during their life time, and produce magnetic vortices analyzed in this paper.

The presence of the non-ideal Hall and ambipolar effects introduces new scales into the system. The typical temporal and spatial scales associated to these effects can be evaluated from the values of the coefficients in Eqs. (\ref{eq:etaa}), as in Khomenko et al. \cite{Khomenko+etal2014b},
\begin{equation}
\tau_A=\eta_A/v_A^2; \,\,\, \tau_H=\eta_H/v_A^2; \,\,\, L_A=\eta_A/v_A; \,\,\, L_H=\eta_H/v_A, 
\end{equation}
where $v_A$ is the Alfv\'en speed. The $\tau_H$ scale is the same quantity as the Hall parameter $\epsilon=\omega/(\omega_{ci}\xi_i)$, used in Paper II. It relates the wave frequency to the ion-cyclotron frequency, and it is inversely proportional to the ion fraction, $\xi_i$. The scaling with  $\xi_i$ increases the importance of the Hall effect toward the lower frequencies in the partially ionized medium.  As was discussed in Khomenko et al. \cite{Khomenko+etal2014b}, the spatial scale associated to the Hall effect, $L_H$, and the temporal scale of the ambipolar effect, $\tau_A$, are independent of the magnetic field strength, while $\tau_H\sim B^{-1}$, and $L_A\sim B$.  Figure \ref{fig:scales} shows two dimensional histograms of the scales computed from the whole time series of snapshots of the \ambihall\ run. The scales are extremely short below the photosphere, and are not shown in the plot. Non-ideal effects play no role at scales covered by the simulations in the sub-photospheric layers. The $\tau_A$ reaches the values larger than 0.01 s above $\sim$1 Mm, while $\tau_H$ does so almost over the whole photosphere and above. These temporal scales are well resolved by the simulations. The spatial scales, $L_A$ and $L_H$, both reach the values of $10^3-10^4$ m at the upper part of the domain. The grid size of our simulations is 20 km, so the scales of the ambipolar and Hall effects at some locations reach, in order of magnitude, the grid size.

\section{Results}

We first show the statistical results of the comparison of the power spectra of vorticity and other related quantities over the 2 hours of simulations. Then, modification of the average magnetic structure of the atmosphere due to the non-ideal effects is discussed. Finally we consider an example of evolution of a particularly long-living magnetic flux tube in the \ambihall\ simulation, which we were able to trace during entire time-span of two hours.

\begin{figure}  
\begin{center}
\includegraphics[width = 13cm]{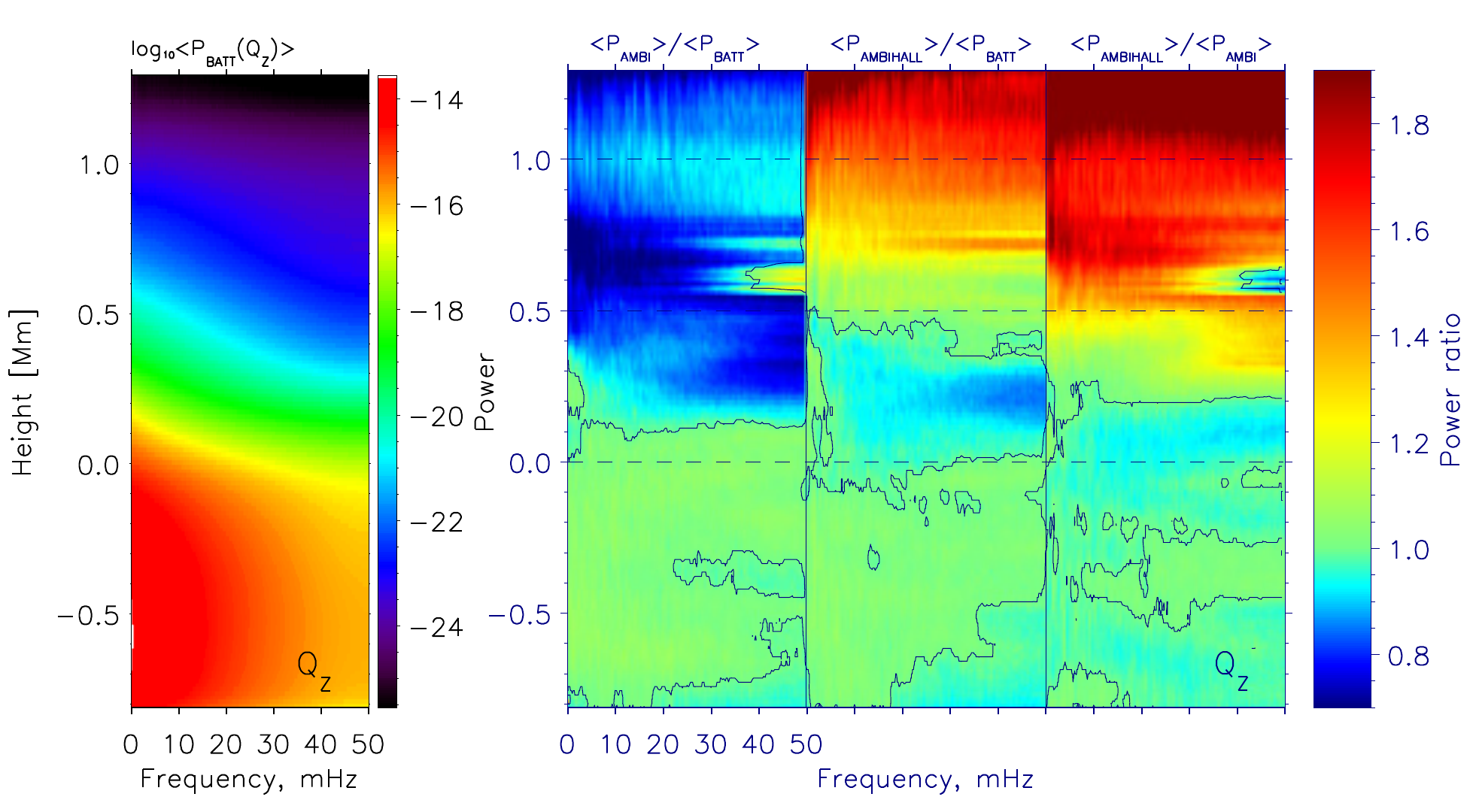}
\includegraphics[width = 13cm]{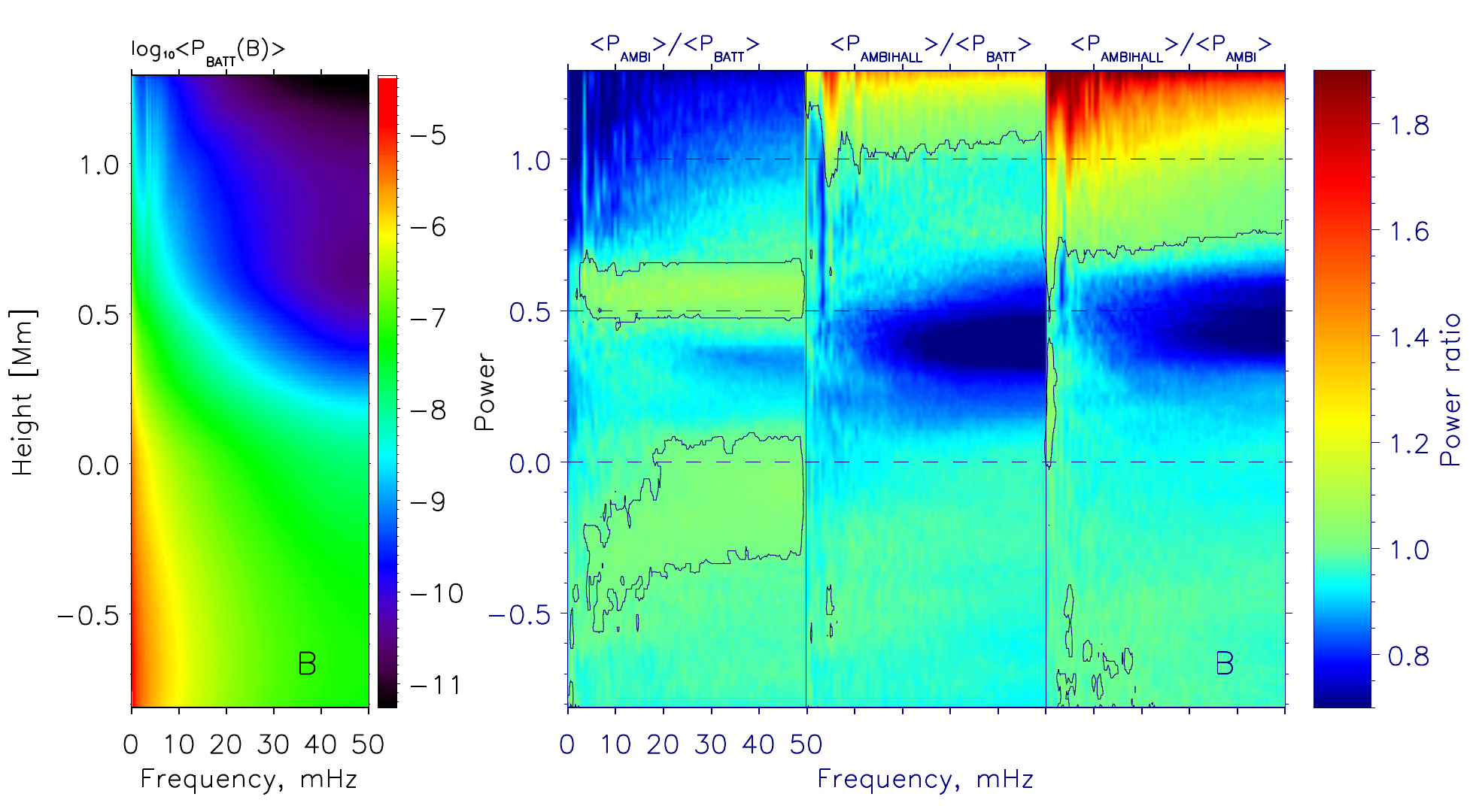}
\end{center}
\caption{Upper left: power spectra of $Q_z$ (defined by Eq. \ref{eq:qcrit}) scaled with the factor of $\rho^2$ for better visualization,  as a function of height and frequency, for \batt\ run.  The power is shown in log$_{10}$ SI units. Upper right: maps of the power ratio, $F(z,\nu)$  for the \ambi\ and \batt, \ambihall\ and \batt, and \ambihall\ and \ambi\ runs. Contours mark the locations where the power ratio is unity. Bottom panels: same for the magnetic field strength, $B$.}
\label{fig:vorticity_power}
\end{figure}

\subsection{Statistical comparison of the simulations}

Turbulent convective motions in realistic simulations generate in a natural way a spectrum of different waves. Here we are interested in incompressible perturbations, associated with Alfv\'en waves, and in the vorticity. In Paper I and II we split the compressible and incompressible perturbations by computing the divergence and the curl of the velocity field parallel to the magnetic field lines. This separation works well independently of plasma $\beta$, and the incompressible perturbation can be associated with Alfv\'en waves \citep{Cally2017, Przybylski+etal2017, Khomenko+etal2018, 2020A&A...XXXX..XXX}. The behavior of Alfv\'en waves and the vorticity is intimately related \cite{Shelyag2011, Shelyag2013}. In order to study the influence of the ambipolar and Hall effects on vorticity, here we computed the proxy for vorticity from the three simulation series using the Q-criterion,

\begin{equation}\label{eq:qcrit}
Q=\frac{1}{2}(|\Omega|^2 - |S|^2) > 0
\end{equation} 
where
\begin{equation}
\Omega_{ij}=\frac{\partial u_i}{\partial x_j} - \frac{\partial u_j}{\partial x_i}; \,\,\, S_{ij}=\frac{\partial u_i}{\partial x_j} + \frac{\partial u_j}{\partial x_i}
\end{equation}
are the vorticity tensor and the rate-of-strain tensor. Taking the velocity field from simulations, we have computed $\Omega_{xy}$ and $S_{xy}$ for the horizontal velocity field components, i.e. the vorticity and the strain in the vertical $z$ direction, and the corresponding $Q_z$. Then, $Q_z$ was Fourier-transformed in time for every point of the domain and frequency-height power maps were constructed for the \batt, \ambi, and \ambihall\ simulations, as

\begin{equation}\label{eq:P}
P(z,\nu)_R= \left<\left|\mathrm{FFT}\left(Q_z \right)_\mathrm{R}\right|^2\right>_{x,y},
\end{equation}
where $R$ represents the run, and the averaging of the power is performed in both horizontal directions. Then the power ratio maps were produced as $F(z,\nu)=P(z,\nu)_{R1}/P(z,\nu)_{R2}$, where $R1, R2$ stand for either of  \batt, \ambi, and \ambihall.

Figure \ref{fig:vorticity_power} (upper panel) presents the results of this calculation. The spectrum of $Q_z$ presents a smooth decrease of power toward high frequencies with no evidence of oscillatory peaks. Its frequency dependence, with power decreasing with frequency, is typical for a spectrum of velocity perturbations generated by convection \cite{Nordlund+etal2009}. The decrease of the power in height owns to multiple reasons, as the existence of cut-off frequencies for low-frequency perturbations, a particular structure of the magnetic field with horizontal structures dominating over the vertical ones at some height, the non-adiabaticity of perturbations due to radiation, etc. The power maps ratio shown at the right reveal considerable differences in the behavior of vorticity between the three simulations. Ambipolar diffusion creates a deficit of vorticity above 200 km. The lower ``absorption'' band between 200 and 500 km has a maximum toward high frequencies. Since the action of the non-ideal effects, such as ambipolar diffusion, is supposed to increase toward small scales, the observed behavior must be related with the dissipation of small scale vorticity motions. Unlike that, the absorption bands at heights above 500 km peak at lower frequencies. This fact is indicative for the overall reorganization of the magnetic structure being responsible for  the lack of the vorticity in the \ambi\ simulations.

The simulations with the Hall effect show an excess of vorticity above 500 km both relative to the \batt\ and \ambi\ simulations up to a factor 2. The dependence of this excess on frequency shows a weak maximum at low frequencies, and therefore is indicative of global magnetic structure changes. Comparing Figure \ref{fig:vorticity_power} above with Figure 3 from Paper II, it becomes evident that $Q_z$ is affected by the ambipolar and Hall effects in a similar way as the incompressible field-aligned perturbation, $\mathrm{f}_{\rm alf}= \hat{e}_\parallel \,\mathbf{\cdot \nabla \times} \mathbf{v}$. Therefore, the ambipolar effect provides a mechanism to dissipate vortices, i.e. to extract energy from the vortices and convert it into thermal energy of the plasma at chromospheric heights. On the contrary, the Hall effect results in a large power of $Q_z$ in the same height range, i.e. in stronger vortices. Since $Q_z$ is quadratic in velocity, the amplitude of the power depression or excess is stronger than for $\mathrm{f}_{\rm alf}$. The joint action of the ambipolar and Hall effects, apparently, provides a way to transport stronger vorticity to the chromosphere and to efficiently dissipate it there. 

The bottom panel of Figure \ref{fig:vorticity_power} shows the power maps computed following Eq. \ref{eq:P}, but for the modulus of magnetic field, $B$. The appearance of the power ratio maps is similar to the case of $Q_z$, with a lack of power caused by the ambipolar diffusion and an excess of power caused by the Hall effect in upper layers. Again, the maximum of the power depression or excess falls at low frequencies. Less magnetic field reaches the upper layers when the ambipolar diffusion is acting, while the Hall effect results in stronger magnetic fields. Nevertheless, there is an additional power depression band present in the ratios \ambihall\ to \batt\ and \ambihall\ to \ambi\ around 500 km which peaks toward high frequencies. High-frequency magnetic field strength fluctuations are less powerful when the Hall effect is acting. 

\begin{figure}  
\begin{center}
\includegraphics[width = 9cm]{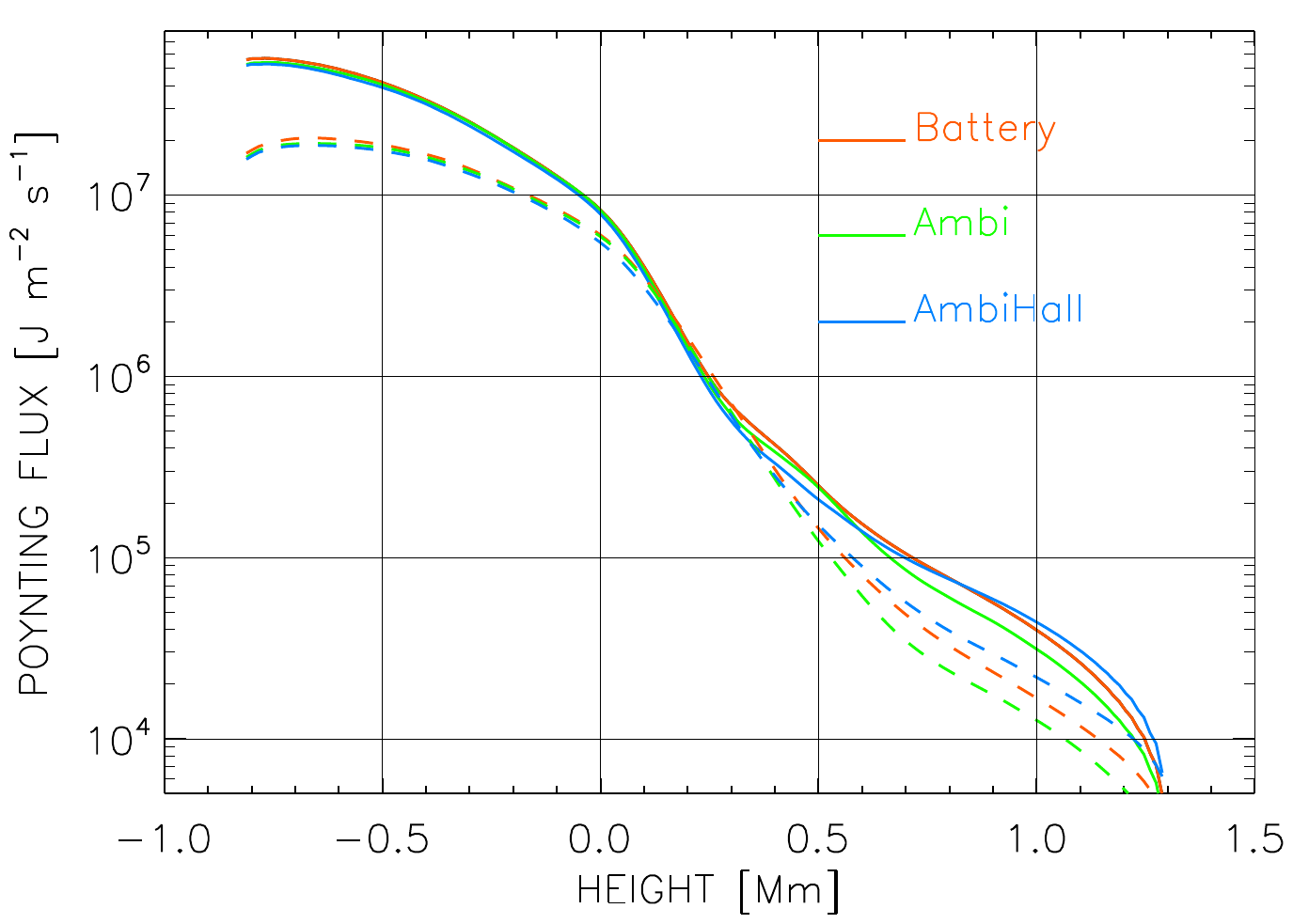}
\end{center}
\caption{Solid lines: space-time average amplitude of $S_{\rm EM}^{\rm ideal, emer}$ component of the Poynting flux as a function of height. The results for \batt, \ambi, and \ambihall\ runs are indicated by the red, green and blue colors, respectively. Dashed lines: space-time average amplitude of   $S_{\rm EM}^{\rm ideal, shear}$ component of the Poynting flux. }
\label{fig:poynting_power}
\end{figure}

Next we investigate how the electro-magnetic Poynting flux, propagating through the domain, is affected by the non-ideal effects. For that we have split the $z$ component of the ideal part of the Poynting flux vector,

\begin{equation}
\mathbf{S}^{\rm ideal}_{\rm EM}=-\frac{(\mathbf{v} \times \mathbf{B}) \times \mathbf{B}}{\mu_0},
\end{equation}
into the ``shear'' and ``emerging'' parts, following Shelyag et al.  \cite{Shelyag2012},

\begin{eqnarray} \label{eq:poynting_ideal}
S_{\rm EM}^{\rm ideal, emer}&=&v_z(B_x^2 + B_y^2)/\mu_0; \\ 
S_{\rm EM}^{\rm ideal, shear}&=&-B_z(B_x v_x + B_y v_y)/\mu_0. 
\end{eqnarray}

\noindent The ``shear'' part corresponds to horizontal motions along vertical flux tubes, while the ``emerging'' part corresponds to horizontal magnetic field perturbations transported by vertical plasma motions. Figure \ref{fig:poynting_power} shows the averaged amplitude of $S_{\rm EM}^{\rm ideal, emer}$ (solid lines) and $S_{\rm EM}^{\rm ideal, shear}$ (dashed lines) over the horizontal coordinates and time for the three runs. The shear component of $S_{\rm EM}^{\rm ideal}$ appears to be larger than the emerging one both below and above the photosphere in all three simulations. At heights between 100 and 400 km the amplitudes of both components coincide. Therefore, horizontal motions along vertical flux tubes produce on average less ideal Poynting flux in our simulations (i.e. $S_{\rm EM}^{\rm ideal, shear} < S_{\rm EM}^{\rm ideal, emer}$). This conclusion is similar to those reached earlier in simulations by Steiner et al. \cite{Steiner2008} who compared the runs with initially vertical implanted field and horizontal advected field. However, in similar simulations by Shelyag et al. \cite{Shelyag2012} only 5\% of the total Poynting flux was generated by advection of horizontal magnetic field by vertical motions, and the dominant component of the flux was produced by the torsional motions in vertical magnetic flux tubes. The magnetic field configuration in our simulations is different from both, Steiner et al. \cite{Steiner2008} and from Shelyag et al. \cite{Shelyag2012}. As was shown in Paper I, the magnetic field lines form low-lying loops and a carpet of horizontal fields in the upper photosphere, with only few vertical flux-tube like concentrations. Therefore, it is not surprising that in our simulations the $S_{\rm EM}^{\rm ideal, shear}$ component is less important than the $S_{\rm EM}^{\rm ideal, emer}$ one.

Figure \ref{fig:poynting_power} reveals that both Poynting flux components are affected by non-ideal effects in a similar way, i.e. the ambipolar effect produces a flux reduction, and the Hall effect produces its excess. However, the amplitude of both, flux reduction and excess, relative to the \batt\ simulation is larger for the $S_{\rm EM}^{\rm ideal, shear}$ component (dashed lines). The total Poynting flux vector in the presence of non-ideal effects follows the expression:

\begin{equation}\label{eq:pf}
\mathbf{S}_{\rm EM}= \mathbf{S}^{\rm ideal}_{\rm EM} - \eta_H  \frac{|B|\mathbf{J}_\perp}{\mu_0} -\frac{\mathbf{B} \times\left(\eta_A \mathbf{J}_\perp \right)}{\mu_0}-\frac{\nabla p_\mathrm{e} \times \mathbf{B}}{en_e\mu_0},
\end{equation}
where $\mathbf{J}_{\perp} = - [(\mathbf{J} \times \mathbf{B})]\times\mathbf{B}/|B|^2$ is the current perpendicular to the magnetic field. 
The non-ideal terms present at the right hand side are the Hall, ambipolar and battery terms. We do not consider the battery term further here due to its smallness. 

The $z$ component of the non-ideal terms can be cast in the following form,

\begin{eqnarray}  \label{eq:poynting_nonideal}
\mathbf{S}_{\rm EM}^{\rm Hall}&=&v_z^{\rm Hall}(B_x^2 + B_y^2)/\mu_0 - B_z(B_x v_x^{\rm Hall} + B_y v_y^{\rm Hall})/\mu_0, \\ \nonumber
\mathbf{S}_{\rm EM}^{\rm Amb}&=&v_z^{\rm Amb}  (B_x^2 + B_y^2) /\mu_0 + B_z(B_z v_z^{\rm Amb})/\mu_0 ,
\end{eqnarray}
where, following Martinez-Sykora et al. \cite{MartinezSykora+etal2012}, we defined Hall and Ambipolar velocities, $v^{\rm Hall}=\eta_H\mathbf{J}/|B|$ and $v^{\rm Amb}=\eta_A[\mathbf{J}\times \mathbf{B}]/B^2$. By comparing Eqs. \ref{eq:poynting_ideal} and  \ref{eq:poynting_nonideal}, one can distinguish the ``shear'' and ``emerging'' components of  $\mathbf{S}_{\rm EM}^{\rm Hall}$ and $\mathbf{S}_{\rm EM}^{\rm Amb}$. The term $\mathbf{S}_{\rm EM}^{\rm Hall}$ closely resembles the behavior of $\mathbf{S}_{\rm EM}^{\rm ideal}$, where the vertical velocity $v^{\rm Hall}$ is being caused by vertical currents, i.e. by vertical plasma motions. This is not so in the case of $\mathbf{S}_{\rm EM}^{\rm Amb}$. In the latter case, the vertical velocity, $v^{\rm Amb}$, is produced by the cross-product of $\mathbf{J}$ and $\mathbf{B}$, and is proportional to the horizontal components of currents and magnetic field, $(J_xB_y-J_yB_x)$. The same difference also applies to the second, ``shear'' term in the expression for $\mathbf{S}_{\rm EM}^{\rm Amb}$. 

According to the divergence theorem, the time evolution of the total energy integrated in volume
\begin{equation}
\frac{\partial }{\partial t}\int_V e_{\rm tot}dV= -\vec{\nabla}\vec{S}dV=-\int_s\vec{S}d\vec{s}
\end{equation}
is conserved if no flux $\vec{S}$ enters or exits the boundaries of the domain. In a stationary situation, the total flux through the plasma volume should be conserved. In an ideal plasma, the electro-magnetic flux will have only one contribution, $\mathbf{S}_{\rm EM}^{\rm ideal}$. However, when the non-ideal terms are present in Eq. \ref{eq:pf}, the amount of $\mathbf{S}_{\rm EM}^{\rm ideal}$ should vary because the total $\mathbf{S}_{\rm EM}$ is to be maintained. The Hall term is non-dissipative and has no contribution to the total energy equation, therefore the Hall term only redistributes the electro-magnetic energy, similar to what the advection term does. This redistribution causes an increase of $\mathbf{S}^{\rm ideal}_{\rm EM}$. The ambipolar effect is dissipative, and causes a decrease of the ideal Poynting flux.

\begin{figure}  
\begin{center}
\includegraphics[width = 11cm]{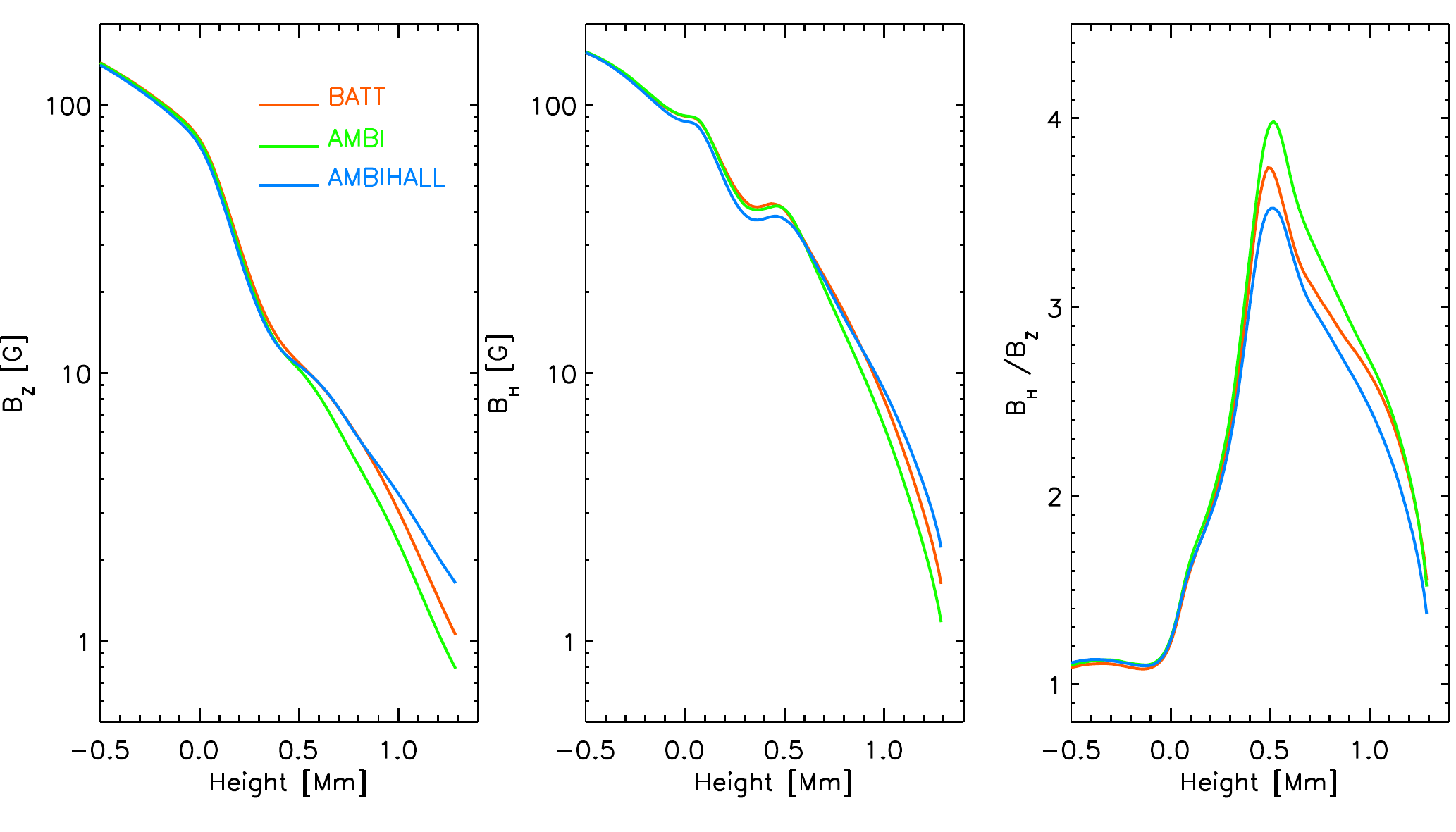}
\end{center}
\caption{Left: horizontally and time-average vertical component of the magnetic field, $B_z$, in the \batt\ (red), \ambi\ (green) and \ambihall\ (blue) runs, as a function of height. Middle: same for the horizontal component, $B_H=\sqrt(B_x^2+B_y^2)$. Right: ratio of $B_H/B_z$. }
\label{fig:bzbh}
\end{figure}

\subsection{Average properties of the magnetic field}

Figure \ref{fig:poynting_power} shows that non-ideal effects influence the ``shear'' component of the ideal Poynting flux more, i.e. the one associated with the horizontal motions along vertical flux tubes. Since these are the motions causing vorticity in the magnetized plasma, it is not surprising that $Q_z$ is affected in a similar way. In order to clarify the reasons of this behaviour, we considered the average properties of the magnetic field in all three runs.

The horizontal and temporal average of the vertical and horizontal components of the magnetic field are displayed in Figure \ref{fig:bzbh}. It shows that differences between the three models are present only in the layers above the surface, in agreement with the fact that both Hall and ambipolar effects are negligible in deep layers. This figure confirms what was already seen in Figure \ref{fig:vorticity_power}, bottom panel, i.e. that the magnetic field is weaker in the upper layers of the \ambi\ run, and stronger in the \ambihall\ run. Figure \ref{fig:bzbh}  demonstrates that this is true for both, vertical and horizontal magnetic field components. However, these components are affected to a different extent. The ratio of $B_H/B_z$ (right panel) reveals that the field emerged to the upper layers is more horizontal in the \ambi\ case, and is more vertical in the \ambihall\ case, compared to the \batt\ one. Therefore, \ambihall\ simulation produces more vertical flux tubes. Since these flux tubes are necessary to propagate vortical motions along them, this can explain an excess of the latter in the \ambihall\ model. 

\begin{figure}  
\begin{center}
\includegraphics[width = 14cm]{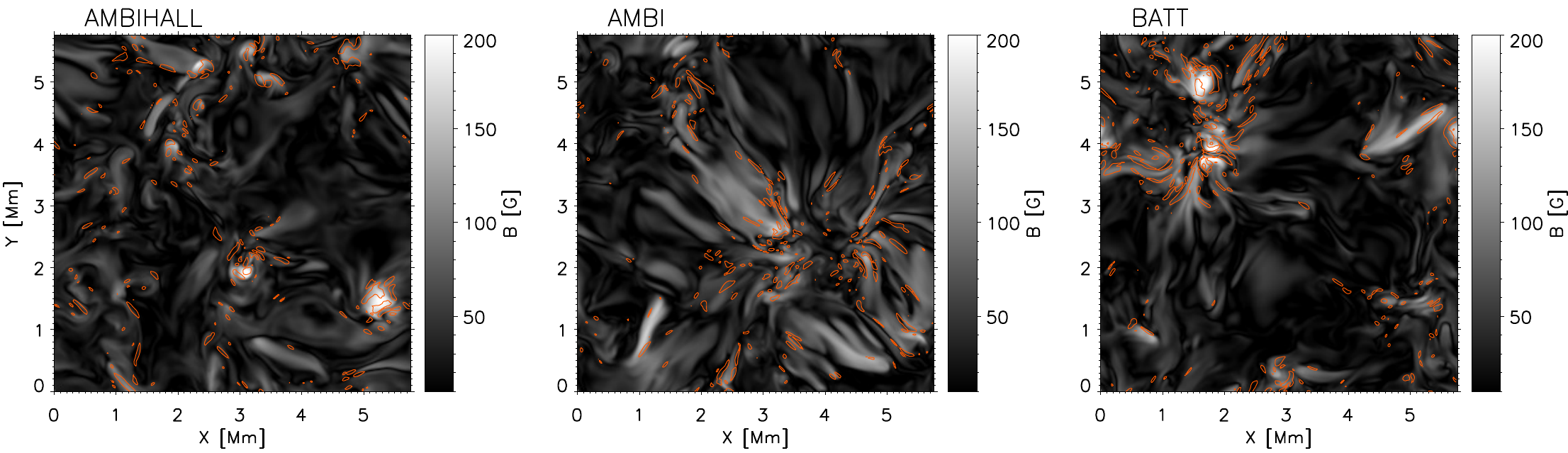}
\end{center}
\caption{Snapshots of the magnetic field strength, $B$, in the \ambihall\ (left), \ambi\ (middle), and \batt\ (right) simulations taken at t=104 min after the start of the analysed series at height 0.56 Mm above the surface. Contours indicate locations of large positive $Q_z$.}
\label{fig:vorticity_b}
\end{figure}

Figure \ref{fig:vorticity_b} illustrates the latter point further. It shows snapshots of the magnetic field strength at an arbitrary time moment, in the upper photosphere. The differences in the structures formed in the three runs are evident by eye. The \ambihall\ run presents few locations with strong and circular magnetic field concentrations, like the one close to $(x,y)=(5.5,1.5)$ Mm. We will analyze this structure closely in the next section. Few other structures of the same kind are present. These flux tubes can be traced almost over the whole duration of the analyzed series. Unlike that, the \ambi\ simulation shows the presence of horizontal fibril-like structures originating from a location around  $(x,y)=(4,2)$ Mm. Again, the formation of such structures is typical in this simulation. The vertical structures are very short lived, and we have not been able to trace any of them for more than several minutes. The \batt\ case shows an intermediate situation between the other two. 

\begin{figure*}  
\begin{center}
\includegraphics[width = 10cm]{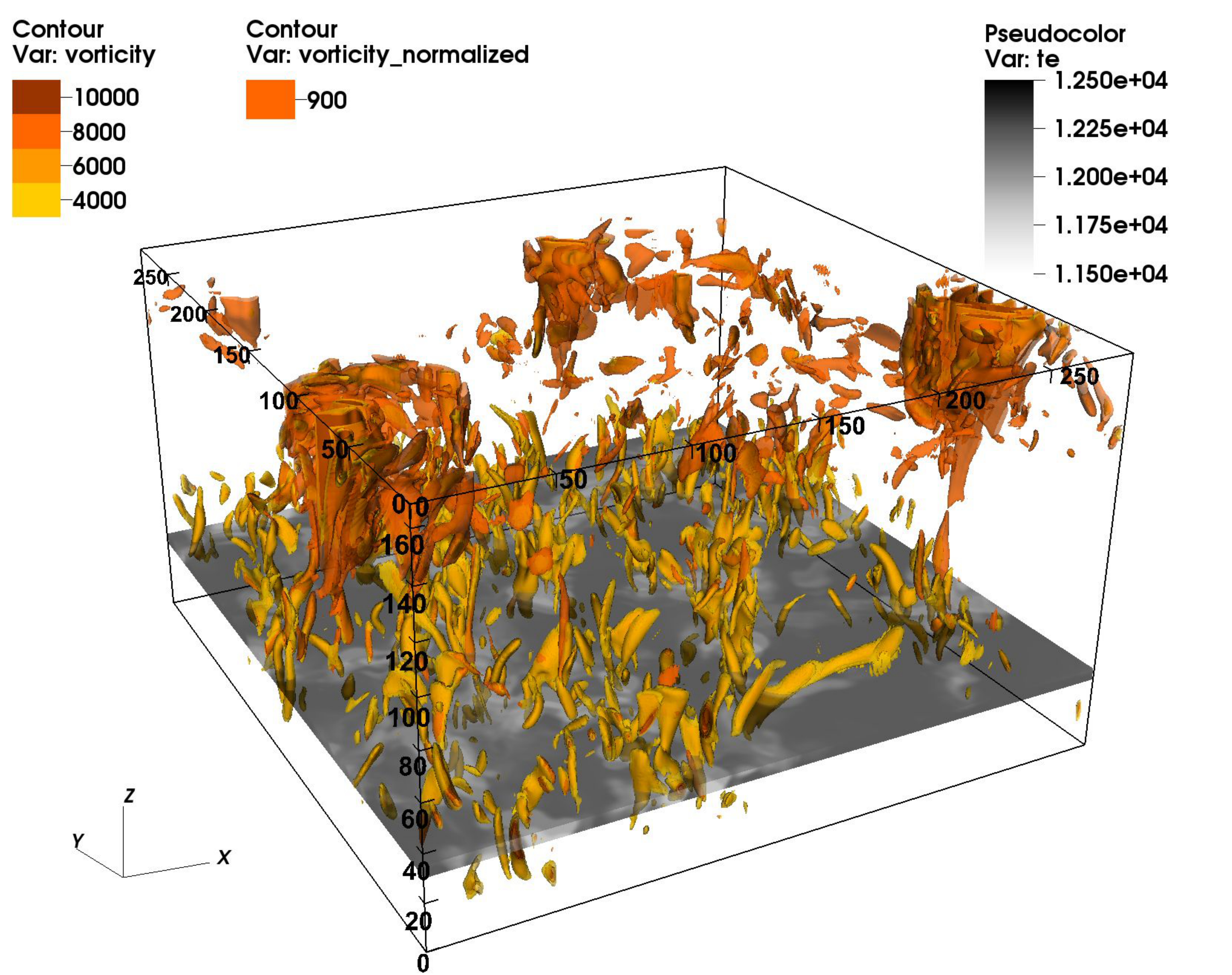}
\includegraphics[width = 10cm]{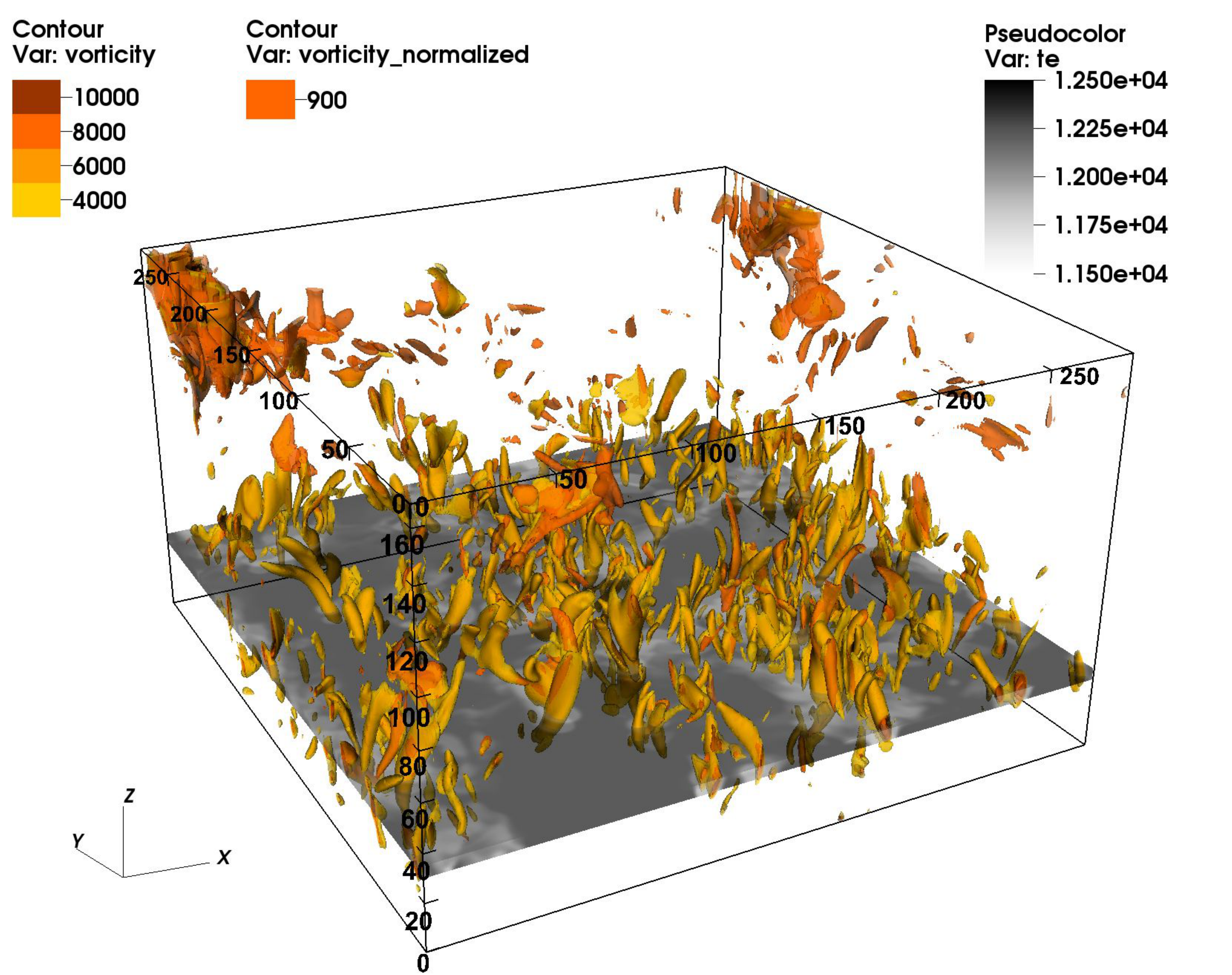}
\end{center}
\caption{3D rendering of the $Q_z$ in the \ambihall\ (upper panel) and \ambi\ simulations. The snapshots are taken at exactly the same time, about 1.5 hours after the Hall effect was introduced in the AMBIHALL model. Bottom image gives temperature at about -0.5 Mm below the photosphere. Yellow contours show locations of positive $Q_z$ above a certain threshold. Orange contours are locations of $Q_z$ normalized to plasma $\beta$, to highlight the location of magnetic vortices. }
\label{fig:vorticity_movie}
\end{figure*}

As has been shown in several studies \cite{Shelyag2011, Moll2012, Wedemeyer2012, Kitiashvili2013, Tziotziou2018, Shetye2019},  magnetic field concentrations serve as funnels for the propagation of the vorticity to the upper layers. A similar behavior is also observed in our simulations. The connection between layers through the magnetic channels is visualized in 3D rendering in Figure \ref{fig:vorticity_movie}. This figure compares two snapshots of the \ambi\ (bottom) and \ambihall\ (top) taken at an arbitrary time moment. Yellow contours follow the locations of positive $Q_z$, irregardless of the magnetic field. It can be seen that vortex structures are present in both simulations in large amounts. They are most visible in the lower part of the domain, located below the surface, and are routed in intergranular lanes. In the sub-surface layers, the hydrodynamic forces are dominant over the magnetic forces,  and the vortices are generated by the baroclinic term in the vorticity equation, as has been shown, for example, by Shelyag et al. \cite{Shelyag2011}. In the magnetic case, most of the vorticity is generated through the term containing the curl of the magnetic tension force \citep{Shelyag2011}.

The vorticity equation in the presence of magnetic field has the following form,

\begin{equation} \label{eq:vorticity-manipulated}
\frac{\partial\vec{\mathbf{\omega}}}{\partial t}  =  \vec{\nabla}\times (\mathbf{v}\times\vec{\mathbf{\omega}}) - \frac{\vec{\nabla}p \times \vec{\nabla}\rho}{\rho^2} + \vec{\nabla}\times\left[\frac{\mathbf{J}\times\mathbf{B}}{\rho}\right],
\end{equation}
where $\vec{\mathbf{\omega}}$ is the vorticity vector field. This equation can be compared to the generalized induction equation,

\begin{eqnarray} \label{eq:induction-manipulated}
\frac{\partial\mathbf{B}}{\partial t}  =  \vec{\nabla}\times (\mathbf{v}\times\mathbf{B}) + \frac{\vec{\nabla}p \times \vec{\nabla}\rho}{\rho^2}\left(\frac{\mu m_p}{e}\right) - \vec{\nabla}\times\left[\frac{\mathbf{J}\times\mathbf{B}}{\rho}\right]\frac{\mu_i m_p}{e\xi_i} +  \vec{\nabla}\times (\eta_A\mathbf{J_{\perp}}),
\end{eqnarray}
where $\mu$ is total mean molecular weight,  $\mu_i$ is the mean molecular weight corresponding only to the ionized fluid, and $m_p$ is proton mass. For the induction equation, it was assumed that the ion fraction $\xi_i=\rho_i/\rho$ and the mean molecular weights are not varying in space. 

It can be seen that the Lorentz force-related term in the vorticity equation, Eq. \ref{eq:vorticity-manipulated}, proportional to $\mathbf{J}\times\mathbf{B}$, is equivalent to the Hall effect-related term in the induction equation, Eq. \ref{eq:induction-manipulated}. The term in the induction equation is multiplied by a coefficient that is inversely proportional to the ion fraction, $\xi_i$. Since the ion fraction in the solar atmosphere can be as low as $10^{-4}$, this  increases the relative importance of the Hall effect, compared to the case of the fully ionised medium. Due to the scaling with $\xi_i$, in partially ionized plasmas the action of the Hall effect extends to lower frequencies  \cite{Pandey+Wardle2008, Pandey2008, Cheung+Cameron2012, Cally+Khomenko2015}.  In current simulations, as shown in Paper II,  the Hall parameter $\epsilon=\omega/(\omega_{ci}\xi_i)$ (or the temporal scale $\tau_H$, see Figure \ref{fig:scales}), reaches values as high as 0.01-1 at heights above 200 km. Therefore, the Hall effect by itself can cause a non-negligible contribution to vorticity generation.
 
In order to distinguish between the magnetic and non-magnetic vortices, we show in Figure \ref{fig:vorticity_movie} the contours of $Q_z$ normalized to the plasma $\beta$ (orange). The magnetic field in our simulations is dynamically weak and the plasma beta achieves values around 1-100 in the upper part of the domain, with only several locations where it drops below 1. The normalization to $\beta$ highlights locations with low beta, i.e. magnetic vortices. It can be seen in the figure that \ambihall\ simulation shows a few locations with vortex tubes connecting the whole domain in the vertical direction. These vortices are related to the vertical magnetic structures present in Figure \ref{fig:vorticity_b} (left), and accompany these structures during their entire life time. No such strong vortex structures can be traced in the \ambi\ simulation. 

\begin{figure*}  
\begin{center}
\includegraphics[width = 12cm]{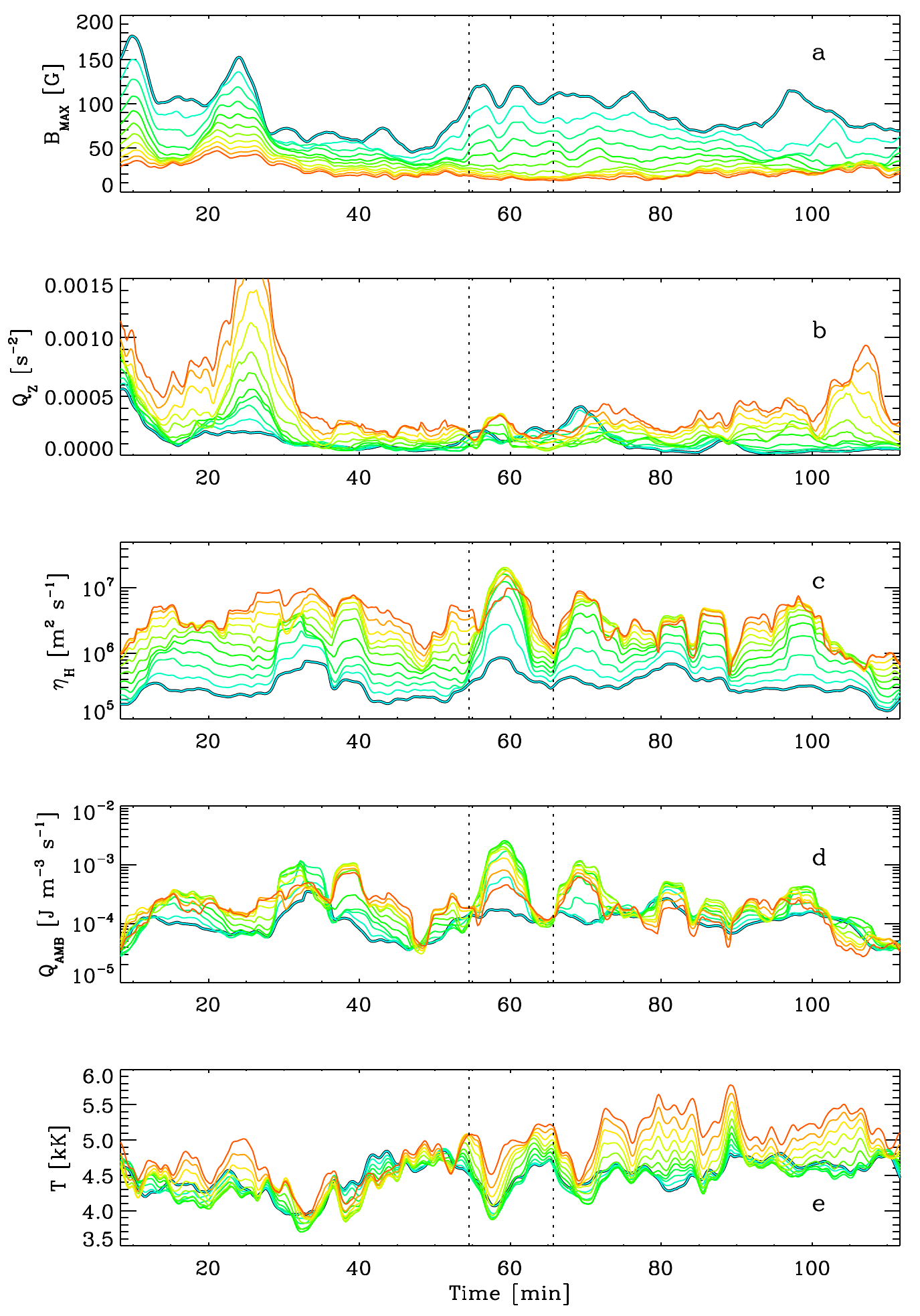}
\end{center}
\caption{Panels from top to bottom: evolution of the maximum magnetic field strength, average positive vorticity $Q_z$, Hall coefficient $\eta_H$, ambipolar heating
$Q_{\rm AMB}$ and temperature as a function of time for an isolated long-living vertical magnetic feature in the \ambihall\ simulation.  The curves from blue to red are for progressively increasing heights from 560 Mm to 980 Mm in steps of 42 km. The variations are smoothed over 200 sec in time to remove 3 minute oscillations and highlight long-term changes. Dotted lines mark the time intervals for the snapshots displayed in Fig. \ref{fig:vorticity_snapshots}.}
\label{fig:vorticity_vars}
\end{figure*}

\begin{figure*}  
\begin{center}
\includegraphics[width = 13cm]{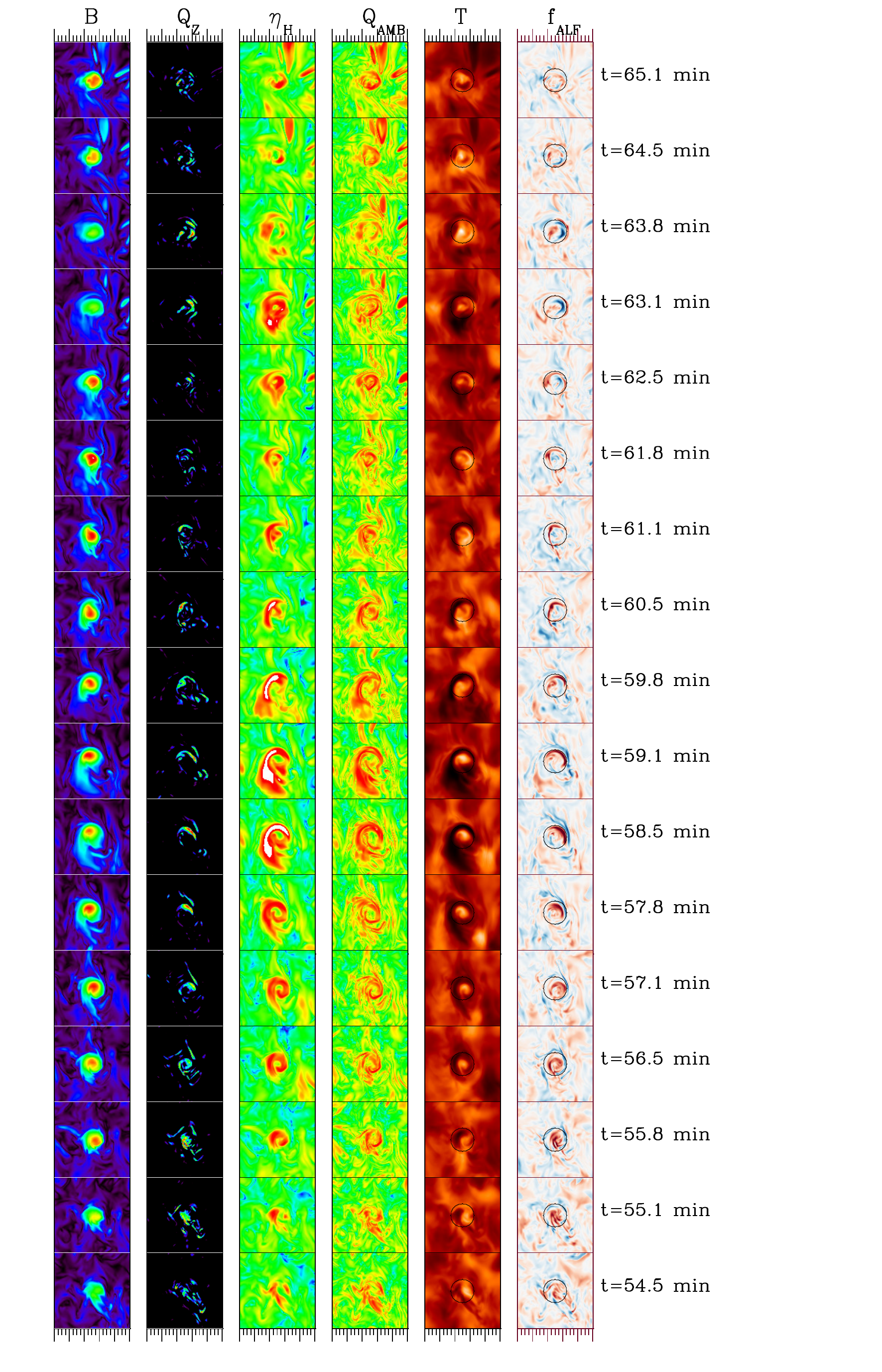}
\end{center}
\caption{Close-up at the evolution of an isolated long-living vertical magnetic feature in the \ambihall\ simulation. Panels from left to right show magnetic field strength, vorticity $Q_z$, Hall coefficient $\eta_H$, ambipolar heating $Q_{\rm AMB}$, temperature, and $f_{\rm alf}$ at height 0.56 Mm in an area of $2\times2$ Mm centered at the feature. Snapshots are shown every 40 seconds, time increases from the bottom to the top, spanning 640 sec in total. Circles at the last two panels highlight the location of the feature.}
\label{fig:vorticity_snapshots}
\end{figure*}

\subsection{Close look at a particular vortex}

In order to clarify the formation of the long-lived vertical flux tubes in the \ambihall\ simulation we have traced one particular magnetic field structure during the 2 hours of the simulations. This structure is visible around the location $(x,y)=(5.5, 1.5)$ Mm in Figure \ref{fig:vorticity_b} (left). We developed a semi-automatic procedure to detect structures with magnetic field and with positive $Q_z$ above a certain threshold, and applied it to a height of 0.56 Mm. 
The initial threshold for the magnetic field detection was 100 G , and for $Q_z$ 0.0005 s$^{-2}$ at 0.56 Mm height.
Once detected, the structure was followed manually to visually confirm its location, and to identity its center as the location with the strongest magnetic field. We set a circle around the center of the structure with a radius of 16 pixels (320 km, which was approximately the average radius of the structure over time), and computed the average values of several parameters: maximum and mean value of the magnetic field strength $B$; maximum and average positive vorticity $Q_z$; average values of the ambipolar heating $Q_{\rm AMB}=\eta_AJ_\perp^2$, of the Hall coefficient $\eta_H$ and temperature, and the amplitude of the field-aligned incompressible perturbation $f_{\rm alf}$. We observe that all these quantities at the height of 0.56 Mm show an oscillatory behavior with a period of about 200 seconds superposed over long-term variations. 

Figure \ref{fig:vorticity_vars} displays the temporal evolution of the computed quantities over the lifetime of the feature. The 3-minute oscillations produce a noisy appearance of the curves and hinder a study of long-term changes. We remove the 3-minute oscillations by smoothing the curves over 200 s intervals. The maximum field strength at 0.56 Mm (blue thick curve) varies between 50 and 150 G. At 0.98 Mm it drops to about 20 G. There are several time intervals when the field strength maintains its maximum at 0.56 Mm. The longest interval is between 55 and 80 min. Variations in the higher layers are less pronounced. The vorticity also shows several peaks that are nearly co-temporal with those of magnetic field strength. The most evident ones are around minutes 8 and 25, both slightly preceding the peaks in $B$. There is also an increase in vorticity between minutes 55 and 75, co-temporal with the magnetic field increase. The vorticity shows larger amplitudes in the higher layers in agreement with the velocity amplitude increase. 

The Hall coefficient (Figure \ref{fig:vorticity_vars} , panel $c$) shows a broad maximum between minutes 55 and 80 at the \hbox{0.56 Mm} height, that is slightly lagged with respect to those in magnetic field and vorticity. The Hall coefficient at the highest layers reaches its maximum at these times over the whole series. The peaks at earlier times do not show evident correlations with those in the magnetic field. The ambipolar heating follows a similar evolution as the Hall coefficient. This is easy to understand because both depend on the electron number density, which is a function of temperature. The temperatures shows several episodes of increase. A maximum in $Q_{\rm AMB}$ at around 33 min coincides with the minimum of temperature, after which the temperature in all the layers starts increasing until minute 50. Then, at minute 57, a new episode of temperature decrease coincides with the maximum heating rate. After approximately 8 minutes from that, at minute 65, the temperature  rises again. Finally, the heating episode that begins at minute 70 produces a steady temperature increase at the upper layers until the end of the time series. The heating episode at minute 57 is co-temporal with the maximum of $\eta_H$ at 0.98 Mm and with a broad maximum of the magnetic field strength at the lower layers. 

Figure \ref{fig:vorticity_snapshots} offers a closer look at the evolution of the feature by showing snapshots of different quantities centred around the magnetic structure. The time interval covered in Figure \ref{fig:vorticity_snapshots} is indicated by dotted lines in Figure \ref{fig:vorticity_vars}. It covers the episodes right after the magnetic field intensification, the heating event around minute 57 and the subsequent temperature increase at minute 65. The evolution of all the quantities shows a clear signature of a vortex. The magnetic feature visibly rotates around its axis and preserves its circular shape. The vorticity, $Q_z$, shows the presence of a structure inside the feature. At time 55.1 min, the maximum $Q_z$ is concentrated at the center of the flux tube. With time, it reveals a spiral structure expanding around the axis with increasing apertures. It is visible, for example, in the interval between times 56.5 and 59.8, when the radius reaches the maximum. Then the structure shrinks again towards the center. The expansion of the structure in $Q_z$ seems to be accompanied by an intensification of the magnetic field, while its shrinking is followed by the magnetic field decay. The $f_{\rm alf}$ follows a similar evolution as $Q_z$, expanding in spiral fronts, which demonstrates the close relation between both quantities. The structures formed in $\eta_H$ and $Q_{\rm AMB}$ are similar. The spiral-like structure intensifies and expands around the location of maximum magnetic field strength forming a circle around the flux tube. The structure is only slightly visible in temperature at time 54.5 min, but at later moments it forms a bright core and a dark surrounding between times 56.5 and 59.8 min. The cool areas around the hot core of the structure are the locations where $Q_{\rm AMB}$ is acting. The Hall coefficient is also the strongest in the immediate surrounding of the flux tube. Possibly, this enhancement of the Hall effect around the flux tube helps to preserve its identity over such a long time interval.  

The behaviour shown in Figure \ref{fig:vorticity_snapshots} is typical for this flux tube. Several episodes of field intensification and decay, spiral motions and heating events are present over its lifetime. Our initial results  show a tight relationship between the propagation of vorticity, intensification of the magnetic flux tube structure and the enhanced action of the Hall effect and ambipolar heating at the flux tubes walls. We plan to perform a statistical analysis of flux tube formation in our simulations in the future.  

\section{Conclusion}

This work presents an initial study of the influence of partial ionization effects on vorticity in realistic magneto-convection simulations. We studied a 2 hours-long series of snapshots of three simulation runs, performed with/without ambipolar and Hall effects and with all other conditions kept fixed. We reached the following conclusions,
\begin{itemize}
\item  The ambipolar effect acts to decrease the vorticity at the upper layers, while the Hall effect acts in the opposite way, and results in up to twice as much vorticity compared to the case when this effect is not present. 

\item The comparison of the average structures of the magnetic field formed in our simulations reveals that stronger and long-lived vertical magnetic field concentrations are formed in the \ambihall\ case, while the \ambi\ case is dominated by more horizontal fields. These vertical flux tubes serve as funnels that connect all the layers in the simulation domain so that vortical motions can propagate along them to the upper layers. 

\item In all the simulations, the ``shear'' component of the Poynting flux, related to the horizontal motions acting on vertical flux tubes, is smaller than the ``emerging'' one. Nevertheless, the ``shear'' component is more affected by the ambipolar and the Hall effect. The ambipolar effect strongly reduces the ``shear'' component by converting magnetic energy into thermal energy of the plasma. 

\item The Hall contribution to the Poynting flux, $\mathbf{S}_{\rm EM}^{\rm Hall}$, mimics the ideal one, 
unlike the ambipolar Poynting flux contribution, $\mathbf{S}_{\rm EM}^{\rm Amb}$. 

\item By comparing the vorticity and the induction equations we conclude that the Hall effect-related term in the induction equation follows the same behavior as the Lorentz force-related term. This Hall-related term is enhanced by a factor inversely proportional to the ion fraction, and therefore it is significantly larger than in a fully ionized plasma. It can therefore contribute to the generation of vorticity in a magnetized environment, similarly to the Lorentz force related term. 

\item A close look into the evolution of a particularly long-lived structure in the \ambihall\ simulation reveals the presence of an internal spiral-like shape in vorticity $Q_z$ and in the field-aligned incompressible perturbation, $f_{\rm alf}$. The flux tube is intensified while the spiral fronts in vorticity are expanded. At the same time, the flux tube is surrounded by a cool zone where the ambipolar heating and the Hall effects are enhanced. We observe that after several heating events in the immediate surroundings of the flux tube, the temperature rises at chromospheric heights.
\end{itemize}

In conclusion, a first analysis of realistic 3D simulations including non-ideal effects shows that these effects are potential mechanisms for vortex (and Alfv\'en waves) dissipation, but also that they can together increase the efficiency to transport vorticity to the chromosphere. Nevertheless, these conclusions are based on a single set of models, with one particular magnetic field setup (local dynamo), and done with one particular numerical code (\mancha). A major effort will be needed in the future to further advance our understanding of the influence of non-ideal effects in partially ionized solar plasmas into the vorticity. 

Here we analyze in detail only one particularly long-lived magnetic feature. In general terms our results are in agreement with previous works \cite{Shelyag2011, Moll2012, Wedemeyer2012, Kitiashvili2013, Tziotziou2018, Shetye2019}: we observe that vertical magnetic structures act as channels to propagate vorticity to the upper layers. More statistics will be needed to fully understand why the formation of these structures is more frequent with the Hall effect acting. The close analysis of this single structure reveals that the Hall effect is enhanced in the immediate surroundings of the structure. The temperature is lower in the surroundings, which decreases the ionization fraction and leads to higher Hall and ambipolar coefficients in this area at a height about 0.56 Mm. More examples should be studied to find out whether the presence of this enhanced Hall effect helps increasing the lifetime of the tubes, preserving their stability against the convective motions. In Paper I we found that heating at the low layers of the atmosphere is crucial for the efficient transformation of magnetic energy into a temperature increase of the plasma. If the heating happens higher above, as in chromospheric shocks, then an important part of it is spent into ionization, and not on a temperature increase. Therefore, ambipolar dissipation of the vortices at heights of about 600 km, as is seen in Figure \ref{fig:vorticity_snapshots}, can contribute to the chromospheric energy budget. Indeed we observe that after a few heating events happening inside the studied flux tube, the temperature in the chromosphere is increased for a significant amount of time.

It is also interesting to note that the life time of our structure is rather unusual because of its length. Only one such long-living structure was studied in observations by Tziotziou et al. \cite{Tziotziou2018}, but without apparent relation to the magnetic field. A close comparison between both, observations and simulations will be an interesting task to perform in the future. A number of vortex tracking techniques has been recently applied, as e.g. \cite{Kato2017, Giagkiozis2018, Liu2019}, to detect and study vortices in simulations and observations. Application of such techniques will certainly be needed to increase the statistics of the initial study we presented above.


\enlargethispage{20pt}

\dataccess{The data this study is based on are too large to host on public repositories. However, parts of the data can be requested from the corresponding author, who will be happy to discuss ways to access the data.}

\aucontribute{EK analyzed the data, wrote the manuscript,  and supervised running the simulations; other authors participated in the analysis and the discussion of the results and contributed into the code development; PAGM run the simulations as a part of his PhD thesis project.}

\competing{The author(s) declare that they have no competing interests.}

\funding{This work was supported by the Spanish Ministry of Science through the project PGC2018-095832-B-I00. It contributes to the deliverable identified in FP7 European Research Council grant agreement ERC-2017-CoG771310-PI2FA for the project ``Partial Ionization: Two-fluid Approach''. EK thanks ISSI Bern for support for the team ``The Nature and Physics of Vortex Flows in Solar Plasmas''.}

\ack{EK wishes to acknowledge scientific discussions with the Waves in the Lower Solar Atmosphere (WaLSA; www.walsa.team) team. The authors thankfully acknowledge the technical expertise and assistance provided by the Spanish Supercomputing Network (Red Espa\~nola de Supercomputaci\'on), as well as the computer resources used: LaPalma Supercomputer, located at the Instituto de Astrof\'isica de Canarias, and MareNostrum based in Barcelona/Spain.}




\providecommand{\noopsort}[1]{}\providecommand{\singleletter}[1]{#1}%


\begin{thebibliography}{10}

\bibitem{Bonet2008}
{Bonet} JA, {M{\'a}rquez} I, {S{\'a}nchez Almeida} J, {Cabello} I, {Domingo} V.
\newblock {Convectively Driven Vortex Flows in the Sun}.
\newblock ApJ. 2008 Nov;687(2):L131.

\bibitem{Vargas2011}
{Vargas Dom{\'\i}nguez} S, {Palacios} J, {Balmaceda} L, {Cabello} I, {Domingo}
  V.
\newblock {Spatial distribution and statistical properties of small-scale
  convective vortex-like motions in a quiet-Sun region}.
\newblock MNRAS. 2011 Sep;416(1):148--154.

\bibitem{OrozcoSuarez2012}
{Orozco Su{\'a}rez} D, {Asensio Ramos} A, {Trujillo Bueno} J.
\newblock {Evidence for Rotational Motions in the Feet of a Quiescent Solar
  Prominence}.
\newblock ApJ. 2012 Dec;761(2):L25.

\bibitem{Tziotziou2020}
Tziotziou K, Scullion E, Jess D, \mbox{et al }.
\newblock {Vortex flows in the solar atmosphere. Definitions, observations,
  modelling and theory}.
\newblock Space Sci\ Rev. 2020;in preparation.

\bibitem{Wedemeyer2009}
{Wedemeyer-B{\"o}hm} S, {Rouppe van der Voort} L.
\newblock {Small-scale swirl events in the quiet Sun chromosphere}.
\newblock A\&A. 2009 Nov;507(1):L9--L12.

\bibitem{Wedemeyer2012}
{Wedemeyer-B{\"o}hm} S, {Scullion} E, {Steiner} O, {Rouppe van der Voort} L,
  {de La Cruz Rodriguez} J, {Fedun} V, et~al.
\newblock {Magnetic tornadoes as energy channels into the solar corona}.
\newblock Nat. 2012 Jun;486:505--508.

\bibitem{Park+etal2016}
{Park} SH, {Tsiropoula} G, {Kontogiannis} I, {Tziotziou} K, {Scullion} E,
  {Doyle} JG.
\newblock {First simultaneous SST/CRISP and IRIS observations of a small-scale
  quiet Sun vortex}.
\newblock A\&A. 2016 Feb;586:A25.

\bibitem{Yadav2020}
{Yadav} N, {Cameron} RH, {Solanki} SK.
\newblock {Simulations Show that Vortex Flows Could Heat the Chromosphere in
  Solar Plage}.
\newblock ApJ. 2020 May;894(2):L17.

\bibitem{Shetye2019}
{Shetye} J, {Verwichte} E, {Stangalini} M, {Judge} PG, {Doyle} JG, {Arber} T,
  et~al.
\newblock {Multiwavelength High-resolution Observations of Chromospheric Swirls
  in the Quiet Sun}.
\newblock ApJ. 2019 Aug;881(1):83.

\bibitem{Tziotziou2019b}
{Tziotziou} K, {Tsiropoula} G, {Kontogiannis} I.
\newblock {A persistent quiet-Sun small-scale tornado. II. Oscillations}.
\newblock A\&A. 2019 Mar;623:A160.

\bibitem{Morton2013}
{Morton} RJ, {Verth} G, {Fedun} V, {Shelyag} S, {Erd{\'e}lyi} R.
\newblock {Evidence for the Photospheric Excitation of Incompressible
  Chromospheric Waves}.
\newblock ApJ. 2013 May;768(1):17.

\bibitem{Tziotziou2018}
{Tziotziou} K, {Tsiropoula} G, {Kontogiannis} I, {Scullion} E, {Doyle} JG.
\newblock {A persistent quiet-Sun small-scale tornado. I. Characteristics and
  dynamics}.
\newblock A\&A. 2018 Oct;618:A51.

\bibitem{Fedun+etal2011}
{Fedun} V, {Shelyag} S, {Erd{\'e}lyi} R.
\newblock {Numerical Modeling of Footpoint-driven Magneto-acoustic Wave
  Propagation in a Localized Solar Flux Tube}.
\newblock ApJ. 2011 Jan;727(1):17.

\bibitem{Shelyag+etal2016}
{Shelyag} S, {Khomenko} E, {de Vicente} A, {Przybylski} D.
\newblock {Heating of the Partially Ionized Solar Chromosphere by Waves in
  Magnetic Structures}.
\newblock ApJ. 2016 Mar;819:L11.

\bibitem{Snow+etal2018}
{Snow} B, {Fedun} V, {Gent} FA, {Verth} G, {Erd{\'e}lyi} R.
\newblock {Magnetic Shocks and Substructures Excited by Torsional Alfv{\'e}n
  Wave Interactions in Merging Expanding Flux Tubes}.
\newblock ApJ. 2018 Apr;857(2):125.

\bibitem{Chmielewski014}
{Chmielewski} P, {Murawski} K, {Solov'ev} ArA.
\newblock {Numerical simulations of three-dimensional magnetic swirls in a
  solar flux-tube}.
\newblock Research in Astronomy and Astrophysics. 2014 Jul;14(7):855--863.

\bibitem{Zaqarashvili2015}
{Zaqarashvili} TV, {Zhelyazkov} I, {Ofman} L.
\newblock {Stability of Rotating Magnetized Jets in the Solar Atmosphere. I.
  Kelvin-Helmholtz Instability}.
\newblock ApJ. 2015 Nov;813(2):123.

\bibitem{Khomenko+Cally2019}
{Khomenko} E, {Cally} PS.
\newblock {Fast-to-Alfv{\'e}n Mode Conversion and Ambipolar Heating in
  Structured Media. II. Numerical Simulation}.
\newblock ApJ. 2019 Oct;883(2):179.

\bibitem{Nordlund+Stein2001}
{Nordlund} {\AA}, {Stein} RF.
\newblock {Solar Oscillations and Convection. I. Formalism for Radial
  Oscillations}.
\newblock ApJ. 2001 Jan;546:576--584.

\bibitem{Stein+Nordlund2001}
{Stein} RF, {Nordlund} {\AA}.
\newblock {Solar Oscillations and Convection. II. Excitation of Radial
  Oscillations}.
\newblock ApJ. 2001 Jan;546:585--603.

\bibitem{Moll2012}
{Moll} R, {Cameron} RH, {Sch{\"u}ssler} M.
\newblock {Vortices, shocks, and heating in the solar photosphere: effect of a
  magnetic field}.
\newblock A\&A. 2012 May;541:A68.

\bibitem{Kitiashvili2013}
{Kitiashvili} IN, {Kosovichev} AG, {Lele} SK, {Mansour} NN, {Wray} AA.
\newblock {Ubiquitous Solar Eruptions Driven by Magnetized Vortex Tubes}.
\newblock ApJ. 2013 Jun;770:37.

\bibitem{Carlsson2010}
{Carlsson} M, {Hansteen} VH, {Gudiksen} BV.
\newblock {Chromospheric heating and structure as determined from high
  resolution 3D simulations .}
\newblock Memorie della Societa Astronomica Italiana. 2010 Jan;81:582.

\bibitem{Moll2011}
{Moll} R, {Cameron} RH, {Sch{\"u}ssler} M.
\newblock {Vortices in simulations of solar surface convection}.
\newblock A\&A. 2011 Sep;533:A126.

\bibitem{Kitiashvili2012}
{Kitiashvili} IN, {Kosovichev} AG, {Mansour} NN, {Wray} AA.
\newblock {Dynamics of Magnetized Vortex Tubes in the Solar Chromosphere}.
\newblock ApJ. 2012 May;751(1):L21.

\bibitem{Wedemeyer2014}
{Wedemeyer} S, {Steiner} O.
\newblock {On the plasma flow inside magnetic tornadoes on the Sun}.
\newblock PASJ. 2014 Dec;66:S10.

\bibitem{Liu2019}
{Liu} J, {Carlsson} M, {Nelson} CJ, {Erd{\'e}lyi} R.
\newblock {Co-spatial velocity and magnetic swirls in the simulated solar
  photosphere}.
\newblock A\&A. 2019 Dec;632:A97.

\bibitem{Shelyag2011}
{Shelyag} S, {Keys} P, {Mathioudakis} M, {Keenan} FP.
\newblock {Vorticity in the solar photosphere}.
\newblock A\&A. 2011 Feb;526:A5.

\bibitem{Shelyag2012}
{Shelyag} S, {Mathioudakis} M, {Keenan} FP.
\newblock {Mechanisms for MHD Poynting Flux Generation in Simulations of Solar
  Photospheric Magnetoconvection}.
\newblock ApJ. 2012 Jul;753:L22.

\bibitem{Khomenko+Collados2012}
{Khomenko} E, {Collados} M.
\newblock {Heating of the Magnetized Solar Chromosphere by Partial Ionization
  Effects}.
\newblock ApJ. 2012 Mar;747:87.

\bibitem{MartinezSykora+etal2012}
{Mart{\'{\i}}nez-Sykora} J, {De Pontieu} B, {Hansteen} V.
\newblock {Two-dimensional Radiative Magnetohydrodynamic Simulations of the
  Importance of Partial Ionization in the Chromosphere}.
\newblock ApJ. 2012;753:161.

\bibitem{MartinezSykora2017}
{Mart{\'{\i}}nez-Sykora} J, {De Pontieu} B, {Carlsson} M, {Hansteen} VH,
  {N{\'o}brega-Siverio} D, {Gudiksen} BV.
\newblock {Two-dimensional Radiative Magnetohydrodynamic Simulations of Partial
  Ionization in the Chromosphere. II. Dynamics and Energetics of the Low Solar
  Atmosphere}.
\newblock ApJ. 2017 Sep;847:36.

\bibitem{Ballester+etal2018}
{Ballester} JL, {Alexeev} I, {Collados} M, {Downes} T, {Pfaff} RF, {Gilbert} H,
  et~al.
\newblock {Partially Ionized Plasmas in Astrophysics}.
\newblock Space Sci\ Rev. 2018 Mar;214(2):58.

\bibitem{Khomenko+etal2018}
{Khomenko} E, {Vitas} N, {Collados} M, {de Vicente} A.
\newblock {Three-dimensional simulations of solar magneto-convection including
  effects of partial ionization}.
\newblock A\&A. 2018 Oct;618:A87.

\bibitem{Nobrega-Siverio+etal2020}
{N{\'o}brega-Siverio} D, {Moreno-Insertis} F, {Mart{\'\i}nez-Sykora} J,
  {Carlsson} M, {Szydlarski} M.
\newblock {Nonequilibrium ionization and ambipolar diffusion in solar magnetic
  flux emergence processes}.
\newblock A\&A. 2020 Jan;633:A66.

\bibitem{Cheung+Cameron2012}
{Cheung} MCM, {Cameron} RH.
\newblock {Magnetohydrodynamics of the Weakly Ionized Solar Photosphere}.
\newblock ApJ. 2012 May;750:6.

\bibitem{Cally+Khomenko2015}
{Cally} PS, {Khomenko} E.
\newblock {Fast-to-Alfv{\'e}n Mode Conversion Mediated by the Hall Current. I.
  Cold Plasma Model}.
\newblock ApJ. 2015 Dec;814:106.

\bibitem{Gonzalez-Morales+etal2019}
{Gonz{\'a}lez-Morales} PA, {Khomenko} E, {Cally} PS.
\newblock {Fast-to-Alfv{\'e}n Mode Conversion Mediated by Hall Current. II.
  Application to the Solar Atmosphere}.
\newblock ApJ. 2019 Jan;870:94.

\bibitem{Khodachenko+Zaitsev2002}
Khodachenko ML, Zaitsev VV.
\newblock FORMATION OF INTENSIVEMAGNETIC FLUX TUBES IN A CONVERGING FLOWOF
  PARTIALLY IONIZED SOLAR PHOTOSPHERIC PLASMA.
\newblock Astrophysics and Space Science. 2002;279:389.

\bibitem{2020A&A...XXXX..XXX}
{Gonzalez-Morales} PA, {Khomenko} E, {Vitas} N, {Collados} M.
\newblock {Joint action of Hall and ambipolar effects in 3D magneto-convection
  simulations of the quiet Sun. I. Dissipation and generation of waves}.
\newblock arXiv e-prints. 2020 Aug;p. arXiv:2008.10429.

\bibitem{Khomenko+etal2017}
{Khomenko} E, {Vitas} N, {Collados} M, {de Vicente} A.
\newblock {Numerical simulations of quiet Sun magnetic fields seeded by the
  Biermann battery}.
\newblock A\&A. 2017 Aug;604:A66.

\bibitem{Khomenko+Collados2006}
Khomenko E, Collados M.
\newblock Numerical Modeling of Magnetohydrodynamic Wave Propagation and
  Refraction in Sunspots.
\newblock ApJ. 2006;653:739---755.

\bibitem{Felipe+etal2010}
{Felipe} T, {Khomenko} E, {Collados} M.
\newblock {Magneto-acoustic Waves in Sunspots: First Results From a New
  Three-dimensional Nonlinear Magnetohydrodynamic Code}.
\newblock ApJ. 2010 Aug;719:357--377.

\bibitem{Gonzalez-Morales+etal2018}
{Gonz{\'a}lez-Morales} PA, {Khomenko} E, {Downes} TP, {de Vicente} A.
\newblock {MHDSTS: a new explicit numerical scheme for simulations of partially
  ionised solar plasma}.
\newblock A\&A. 2018 Jul;615:A67.

\bibitem{1989GeCoA..53..197A}
{Anders} E, {Grevesse} N.
\newblock {Abundances of the elements - Meteoritic and solar}.
\newblock Geochimica et Cosmochimica Acta. 1989 Jan;53:197--214.

\bibitem{SanchezAlmeida+MartinezGonzales2011}
{S{\'a}nchez Almeida} J, {Mart{\'{\i}}nez Gonz{\'a}lez} M.
\newblock {The Magnetic Fields of the Quiet Sun}.
\newblock In: {Kuhn} JR, {Harrington} DM, {Lin} H, {Berdyugina} SV,
  {Trujillo-Bueno} J, {Keil} SL, et~al., editors. Solar Polarization 6. vol.
  437 of Astronomical Society of the Pacific Conference Series; 2011. p. 451.

\bibitem{TrujilloBueno2004}
{Trujillo Bueno} J, {Shchukina} N, {Asensio Ramos} A.
\newblock {A substantial amount of hidden magnetic energy in the quiet Sun}.
\newblock Nat. 2004 Jul;430:326--329.

\bibitem{Shchukina+TrujilloBueno2011}
{Shchukina} N, {Trujillo Bueno} J.
\newblock {Determining the Magnetization of the Quiet Sun Photosphere from the
  Hanle Effect and Surface Dynamo Simulations}.
\newblock ApJ. 2011 Apr;731(1):L21.

\bibitem{Danilovic2016}
{Danilovic} S, {van Noort} M, {Rempel} M.
\newblock {Internetwork magnetic field as revealed by two-dimensional
  inversions}.
\newblock A\&A. 2016 Sep;593:A93.

\bibitem{Khomenko+etal2014b}
{Khomenko} E, {Collados} M, {D{\'{\i}}az} A, {Vitas} N.
\newblock {Fluid description of multi-component solar partially ionized
  plasma}.
\newblock Physics of Plasmas. 2014 Sep;21(9):092901.

\bibitem{Cally2017}
{Cally} PS.
\newblock {Alfv{\'e}n waves in the structured solar corona}.
\newblock MNRAS. 2017 Apr;466:413--424.

\bibitem{Przybylski+etal2017}
{Przybylski} D, {Shelyag} S, {Khomenko} E, {de Vicente} A.
\newblock {Efficiency of heating of the partially ionized solar chromosphere by
  dissipation of Alfv{\'e}nic waves}.
\newblock ApJ. 2018;submitted.

\bibitem{Shelyag2013}
{Shelyag} S, {Cally} PS, {Reid} A, {Mathioudakis} M.
\newblock {Alfv{\'e}n Waves in Simulations of Solar Photospheric Vortices}.
\newblock ApJ. 2013 Oct;776:L4.

\bibitem{Nordlund+etal2009}
{Nordlund} {\AA}, {Stein} RF, {Asplund} M.
\newblock {Solar Surface Convection}.
\newblock Living Reviews in Solar Physics. 2009;6:2.

\bibitem{Steiner2008}
{Steiner} O, {Rezaei} R, {Schaffenberger} W, {Wedemeyer-B{\"o}hm} S.
\newblock {The Horizontal Internetwork Magnetic Field: Numerical Simulations in
  Comparison to Observations with Hinode}.
\newblock ApJ. 2008 Jun;680:L85.

\bibitem{Pandey+Wardle2008}
Pandey BP, Wardle M.
\newblock Hall magnetohydrodynamics of partially ionized plasmas.
\newblock MNRAS. 2008;385:2269--2278.

\bibitem{Pandey2008}
Pandey BP, Vranjes J, Krishan V.
\newblock Waves in the solar photosphere.
\newblock MNRAS. 2008;386:1635.

\bibitem{Kato2017}
{Kato} Y, {Wedemeyer} S.
\newblock {Vortex flows in the solar chromosphere. I. Automatic detection
  method}.
\newblock A\&A. 2017 May;601:A135.

\bibitem{Giagkiozis2018}
{Giagkiozis} I, {Fedun} V, {Scullion} E, {Jess} DB, {Verth} G.
\newblock {Vortex Flows in the Solar Atmosphere: Automated Identification and
  Statistical Analysis}.
\newblock ApJ. 2018 Dec;869(2):169.

\end{thebibliography}
\end{document}